\documentclass[journal]{IEEEtran}
\ifCLASSINFOpdf
\else
\fi

\usepackage[bookmarks,colorlinks]{hyperref}
\usepackage{enumitem}
\usepackage[linesnumbered,ruled,lined]{algorithm2e}

\usepackage{graphicx}
\usepackage{dcolumn}
\usepackage{bm}

\usepackage[usenames,dvipsnames]{xcolor}

\usepackage{tabularx}
\usepackage{array}

\usepackage{amsmath,amsfonts}

\usepackage{array}
\usepackage[caption=false,font=normalsize,labelfont=sf,textfont=sf]{subfig}
\usepackage{textcomp}
\usepackage{stfloats}
\usepackage{url}
\usepackage{verbatim}
\usepackage{graphicx}
\usepackage{cite}
\usepackage{svg}
\usepackage[noend]{algpseudocode} 
\usepackage{etoolbox}

\usepackage{float}
\usepackage{pgfplots}
\usepackage[bookmarks,colorlinks]{hyperref}
\usepackage[linesnumbered,ruled,lined]{algorithm2e}
\usepackage{enumitem}
\usepackage{algpseudocode}
\usetikzlibrary{shapes.multipart,intersections}
\usepackage{cite}
\usepackage{amsmath,amssymb,amsfonts,amsthm,steinmetz}
\usepackage{graphicx}
\usepackage{mathrsfs}  
\usepackage{textcomp}
\usepackage{acronym}
\usepackage{xcolor}
\usepackage{upgreek,xspace}

\usepackage{tikz}
\usetikzlibrary{calc}
\makeatletter
\newcommand{\gettikzxy}[3]{%
  \tikz@scan@one@point\pgfutil@firstofone#1\relax
  \edef#2{\the\pgf@x}%
  \edef#3{\the\pgf@y}%
}
\makeatother

\usepackage[draft]{todonotes}

\usepackage{esvect}

\usepackage{times}
\usepackage{bm}
\usepackage{amsmath}
\usepackage{amssymb}
\usepackage{stmaryrd}
\usepackage{babel}
\usepackage{graphics, graphicx}
\usepackage{xcolor}
\usepackage{gensymb}
\usepackage{cite}
\usepackage{enumitem}
\usepackage{url}

\begin{document}
%
\title{A physics-compliant \textit{diagonal} representation for wireless channels parametrized by \textit{beyond-diagonal} reconfigurable intelligent surfaces}
%
%
%

\author{Philipp~del~Hougne,~\IEEEmembership{Member,~IEEE}
\thanks{Parts of this work were presented at the 25th IEEE International Workshop on Signal Processing Advances in Wireless Communications (SPAWC) 2024~\cite{diagonal_BD_RIS}.}
\thanks{The author acknowledges funding from the ANR France 2030 program (project ANR-22-PEFT-0005)and the ANR PRCI program (project ANR-22-CE93-0010).}
\thanks{P.~del~Hougne is with Univ Rennes, CNRS, IETR - UMR 6164, F-35000 Rennes, France (e-mail: philipp.del-hougne@univ-rennes.fr).}
}

%
%

\markboth{}%
{Shell \MakeLowercase{\textit{et al.}}: Bare Demo of IEEEtran.cls for IEEE Journals}
%



\maketitle

\begin{abstract}
The parametrization of wireless channels by so-called ``beyond-diagonal reconfigurable intelligent surfaces'' (BD-RIS) is mathematically characterized by a matrix whose off-diagonal entries are partially or fully populated. Physically, this corresponds to tunable coupling mechanisms between the RIS elements that originate from the RIS control circuit. Here, we derive a physics-compliant \textit{diagonal} representation for BD-RIS-parametrized channels. 
We recognize that any RIS control circuit can always be separated into its static parts (SLC) and a set of tunable individual loads (IL). Therefore, a BD-RIS-parametrized channel results from the chain cascade of three systems: \textit{i)} radio environment (RE), \textit{ii)} SLC, and \textit{iii)} IL. RE and SLC are static non-diagonal systems whose cascade K is terminated by the tunable diagonal system IL. This physics-compliant representation in terms of K and IL is directly analogous to that for conventional (``diagonal'') RIS (D-RIS). Therefore, scenarios with BD-RIS can also readily be captured by the physics-compliant coupled-dipole model PhysFad, as we show. In addition, physics-compliant algorithms for system-level optimization with D-RIS can be directly applied to scenarios with BD-RIS. We demonstrate this important implication of our conceptual finding in a case study on end-to-end channel estimation and optimization in a BD-RIS-parametrized rich-scattering environment. Our case study is the first experimentally grounded system-level optimization for BD-RIS: We obtain the characteristics of RE and IL from experimental measurements and a commercial PIN diode, respectively. Altogether, our physics-compliant diagonal representation for BD-RIS enables a paradigm shift in how practitioners in wireless communications and signal processing implement system-level optimizations for BD-RIS because it enables them to directly apply existing physics-compliant D-RIS algorithms.
\end{abstract}


%
\IEEEpeerreviewmaketitle

\section{Introduction}

The properties of the channels used for wireless communications are traditionally imposed by the preexisting radio environment. With the advent of the ``smart radio environment''~\cite{subrt2012intelligent,Liaskos_Visionary_2018,del2019optimally,di2020smart,alexandropoulos2021reconfigurable}, the wireless channels become to some extent controllable because they are deterministically parametrized by so-called reconfigurable intelligent surfaces (RISs) that are placed within the radio environment. An RIS is an array of elements with individually adjustable scattering properties. In practice, an RIS often takes the form of an array of patch antennas whose ports are terminated by individual tunable lumped elements. Both in simplified cascaded channel models as well as in physics-compliant end-to-end channel models, the tunable RIS configuration appears along the diagonal of an otherwise empty matrix (see details in Sec.~\ref{sec_theory} below). For this reason, we refer to such conventional RIS designs as ``diagonal RIS'' (D-RIS) in this paper.

To further increase the RIS-based control over the wireless channels without increasing the number of RIS elements, it has been proposed to go beyond D-RIS with RIS designs whose tunability is mathematically represented by more densely (potentially fully) populated matrices, again both in simplified cascaded models and physics-compliant models (see details in Sec.~\ref{sec_theory} below)~\cite{shen2021modeling,li2023reconfigurable}. Following the terminology put forth in Refs.~\cite{shen2021modeling,li2023reconfigurable}, we refer to such RIS designs as ``beyond-diagonal RIS'' (BD-RIS) in this paper. Physically, these non-zero off-diagonal terms correspond to \textit{controllable coupling effects between RIS elements that originate from the RIS load circuit} as opposed to \textit{uncontrollable mutual coupling via the radio environment due to proximity and reverberation}. Indeed, any physics-compliant model involves potentially significant but uncontrollable interactions between the RIS elements via the radio environment but these are hence qualitatively different from the tunable interactions via the load circuit proposed in BD-RIS.\footnote{The concepts of ``non-local metasurfaces'' and ``non-local metamaterials'' studied in other communities relate to various mechanisms for carefully designed but typically static coupling between meta-atoms~\cite{shastri2023nonlocal,chen2023nonlocal,sol2024covert}. Although in some experimental realizations of reverberation-non-local metasurfaces the meta-atoms are individually programmable~\cite{sol2022meta,sol2023reflectionless,faul2024agile}, these hence do not constitute realizations of the BD-RIS concept.}

To avoid confusion about the terminology, we emphasize that we use the terms ``D-RIS'' and ``BD-RIS'' to characterize the load circuit of the RIS without making any statement about whether the end-to-end channel model neglects or accounts for mutual coupling between RIS elements arising due to their spatial proximity or reverbreation within the radio environment. In other words, a ``D-RIS'' load circuit architecture does \textit{not} imply that a simplified cascaded model neglecting mutual coupling is assumed.

A wide range of designs of the tunable coupling mechanisms between RIS elements in BD-RIS is theoretically conceivable; Refs.~\cite{shen2021modeling,li2023reconfigurable} provide a taxonomy of carefully crafted designs ranging from pair-wise tunable couplings to all-to-all tunable couplings. To date, their experimental realization remains largely an open challenge and, to the best of our knowledge, only tunable load networks allowing for neighbor-to-neighbor couplings were experimentally realized in conceptually equivalent contexts of RFID channel estimation~\cite{denicke2012application} and contactless scattering matrix estimation with a ``virtual vector network analyzer''~\cite{V2NA_2p0}. Incidentally, a simple-to-realize alternative coupling design that has not received any attention to date would be random tunable couplings, implemented, for instance, via a tunable chaotic cavity backing the RIS.

Most research efforts on D-RIS and BD-RIS are primarily focused on system-level optimizations, assuming simplified cascaded channel models. Because the tunability of BD-RIS is represented not by a diagonal but a more densely (potentially fully populated) matrix, the mathematical structure of the channel parametrization appears to differ from that in the D-RIS case, motivating recent efforts to conceive special algorithms for system-level optimization dedicated to BD-RIS architectures. Meanwhile, more and more researchers seek to model and optimize RIS-assisted wireless communications in a way that is electromagnetically consistent, by accounting for various mutual coupling effects. We summarize below these efforts based on multi-port network theory (Sec.~\ref{subsec_background}) or coupled-dipole formalisms (Sec.~\ref{subsec_physfad_background}).

In this paper, we introduce a thus far overlooked and at first sight counter-intuitive perspective on the representation of BD-RIS-parametrized wireless channels. Specifically, we propose a physics-compliant end-to-end representation of BD-RIS-parametrized wireless channels involving a \textit{diagonal} tunable matrix. Our fundamental insight is that irrespective of the BD-RIS load circuit design, the latter can be separated into a static part and a tunable part that typically relies on tunable lumped elements. Therefore, the BD-RIS-parametrized wireless channel is a chain cascade of three systems:\footnote{Mathematically, the physics-compliant evaluation of a chain cascade of multiple systems is generally more complicated than a chain multiplication of three matrices -- see Sec.~\ref{sec_theory}.}
\begin{enumerate}[label=\emph{\roman*)}]
    \item RE: the radio environment,
    \item SLC: the static parts of the load circuit, and
    \item IL: the individual tunable loads.
\end{enumerate}
Only IL is tunable, while RE and SLC are static. Moreover, only IL is characterized by a diagonal scattering matrix, while RE and SLC are generally characterized by more densely populated non-diagonal scattering matrices. Therefore, we propose to group together the two static and non-diagonal systems RE and SLC, singling out the tunable diagonal system IL. Thereby, we obtain a representation of the end-to-end wireless channel involving a static non-diagonal system (cascade of RE and SLC) and a tunable diagonal system (IL). This representation is analogous to the case of D-RIS in which the static non-diagonal system is simply RE and the tunable diagonal system is IL\footnote{While the number of tunable lumped elements equals the number of RIS elements in a D-RIS architecture, this is generally \textit{not} the case in a BD-RIS architecture -- see Sec.~\ref{sec_theory}.}. In contrast to our proposal, the conventional representation of BD-RIS-parametrized channels groups together the static non-diagonal system SLC and tunable diagonal system IL, yielding a representation of the end-to-end channel involving a static non-diagonal system (RE) and a tunable non-diagonal system (cascade of SLC and IL).

The most important implication of our formulation from the perspectives of wireless communications and signal processing is that, in principle, our formulation obviates the need to develop new algorithms specifically dedicated to BD-RIS for system-level optimizations. 
Instead, existing physics-compliant algorithms for D-RIS can be directly applied to BD-RIS by replacing RE in the D-RIS case with the cascade of RE and SLC in the BD-RIS case.

\subsection{Contributions}

Our main conceptual (theoretical) contribution is the introduction of a \textit{physics-compliant diagonal representation of BD-RIS-parametrized channels}. We recognize that the BD-RIS load circuit can be split into SLC and IL, implying that a BD-RIS-parametrized channel arises from the chain cascade of three systems: RE, SLC, IL. Hence, the cascade of RE and SLC, named K, is terminated by the tunable diagonal system IL -- directly analogous to physics-compliant models of D-RIS-parametrized channels. More specific details regarding our conceptual contribution are as follows:
\begin{itemize}
    \item In contrast to the associated conference paper~\cite{diagonal_BD_RIS} based on impedance parameters, we use scattering parameters. Thereby, our formulation is more compact and applies even to canonical BD-RIS load circuit designs for which an impedance or admittance formulation would fail.
    \item  We establish the equivalence between the formulation in terms of scattering parameters and the coupled-dipole-based PhysFad model for D-RIS~\cite{faqiri2022physfad}. Thereby, we clarify how BD-RIS can be treated within PhysFad.
    \item We explain how the scattering parameters of SLC can be identified in various scenarios that can arise in theoretical, numerical and experimental studies. These scenarios range from canonical load circuits based on ideal T or $\pi$ networks with tunable impedances, via  full-wave simulations, to experimental estimations.
    \item We thoroughly explain under what assumptions the widespread cascaded channel model can be derived from the physics-compliant end-to-end model formulated in terms of scattering parameters for a channel parametrized by a D-RIS or BD-RIS.
\end{itemize}

Our main algorithmic contribution is the \textit{experimentally grounded demonstration of physics-compliant end-to-end estimation and optimization of a BD-RIS-parametrized channel using D-RIS algorithms.} We evidence the most important implication of our conceptual finding in the realm of system-level optimization: Existing physics-compliant D-RIS algorithms can be directly applied to BD-RIS. More specific details regarding this contribution are as follows:
\begin{itemize}
    \item We perform end-to-end channel estimation in a rich-scattering BD-RIS-parametrized radio environment using an existing physics-compliant D-RIS algorithm. We highlight the existence and operational irrelevance of parameter ambiguities.
    \item We perform RSSI maximization based on the estimated physics-compliant diagonal end-to-end channel model, achieving optimal performance as confirmed by an exhaustive brute-force search.
    \item This is the first experimentally grounded study on system-level optimization for BD-RIS (all existing papers are purely theoretical). The assumed characteristics of RE are based on experimental measurements in a rich-scattering environment. The assumed characteristics of IL are based on a commercial PIN diode.
\end{itemize}

\subsection{Outline}

The remainder of this paper is organized as follows. 
In Sec.~\ref{sec_theory}, we develop the theory of our physics-compliant diagonal representation for BD-RIS: We start with background on multi-port network models for RIS-parametrized channels (Sec.~\ref{subsec_background}), then we present our key conceptual insight (Sec.~\ref{sub_sec_keyinsight}), and finally we transpose it to the PhysFad model (Sec.~\ref{sec_physfad}). 
In Sec.~\ref{sec_determining_S_LC}, we explain how to identify the scattering properties of SLC analytically for two canonical ideal load circuits, and we discuss how to determine it numerically or experimentally for practical load circuits. 
In Sec.~\ref{sec_e2e_estim}, we demonstrate the direct applicability of existing physics-compliant D-RIS algorithms for end-to-end channel estimation and optimization in an experimentally grounded case study of a BD-RIS-parametrized rich-scattering environment.
In Sec.~\ref{sec_conclusion}, we briefly conclude.

\subsection{Notation}

\begin{itemize}
\item The superscripts $^T$, $^\star$ and $^\dagger$ and  denote the transpose, the conjugate and the transpose conjugate, respectively.
    \item $\mathbf{I}_d$ denotes the $d \times d$ identity matrix.
    \item $\mathbf{0}_d$ denotes the $d \times d$ zero matrix.
    \item $\mathbf{1}_m$ denotes an $m$-element column vector whose entries are all unity.
    \item $\mathrm{diag}(\mathbf{a})$ denotes the matrix whose diagonal entries are those of the vector $\mathbf{a}$ and all other entries are zero.
    \item $\mathrm{vec}(\mathbf{A})$ denotes the vectorization of the matrix $\mathbf{A}$.
    \item $\mathbf{S}^\mathrm{X}$ is the scattering matrix of the multi-port system X.
    \item $\mathbf{S}_\mathcal{FG}^\mathrm{X}$ denotes the block of $\mathbf{S}^\mathrm{X}$ identified by the sets of row [column] indices $\mathcal{F}$ [$\mathcal{G}$].
\end{itemize}

\section{Theory}
\label{sec_theory}

The generic multi-port chain cascade formulation for BD-RIS-parametrized wireless channels developed in this paper is based on multi-port network theory which can equivalently be expressed in terms of scattering, impedance or admittance parameters (background is provided in Sec.~\ref{subsec_background}). Moreover, an equivalent coupled-dipole formalism coined PhysFad~\cite{faqiri2022physfad,prod2023efficient,sol2023experimentally,rabault2023tacit} exists (background is provided in Sec.~\ref{subsec_physfad_background}).
We use scattering parameters as opposed to impedance or admittance parameters for two reasons: first, scattering parameters yield a more compact formulation since the wireless channels are scattering coefficients; second, scattering parameters are suitable to describe canonical examples of BD-RIS load circuits that constitute pathological special cases in which impedance parameters cannot be defined. The general impedance formulation can be found in our associated conference paper~\cite{diagonal_BD_RIS}. A discussion about the relation to PhysFad is provided in Sec.~\ref{sec_physfad}.

Throughout this paper, we assume that the discussed ports and auxiliary ports are monomodal. Otherwise, extensions of the presented theory in terms of generalized scattering parameters~\cite{seguinot1998multimode} would be required. We further make the common and typically applicable assumptions that the considered multi-port networks are linear, passive, time-invariant and reciprocal. We make no assumptions or approximations beyond those stated in this paragraph.

\subsection{Background on Multi-Port Network Models for RIS-Parametrized Channels}
\label{subsec_background}

Building on a decade-old circuit theory of communication~\cite{ivrlavc2010toward,ivrlavc2014multiport}, the literature contains by now many proposed formulations for physics-inspired models of D-RIS-parametrized channels based on multi-port network theory~\cite{shen2021modeling,gradoni_EndtoEnd_2020,tap2022,badheka2023accurate,mursia2023,tapie2023systematic,Akrout_asilomar_2023,del2024minimal,nossek2024,konno2024,nerini2024universal,franek_eucap_2024,viikari2024,abrardo2024}, as well as three formulations for BD-RIS-parametrized channels~\cite{BDRIS_renzo_clerckx_2024,diagonal_BD_RIS,V2NA_2p0}. These numerous formulations differ regarding the approximations and assumptions that they make, especially regarding the nature of the radio environment. In the present paper, we do not make any assumptions beyond those stated in the previous paragraph, ensuring that our formulation is fully physics-compliant. Moreover, we work with the most compact and universal multi-port description of an arbitrarily complex smart radio environment that describes the latter with the scattering matrix $\mathbf{S}^\mathrm{RE}\in \mathbb{C}^{N_\mathrm{RE} \times N_\mathrm{RE}}$ that lumps together all scattering effects, including scattering objects and walls (irrespective of whether they are point-like or not) and structural scattering~\cite{king1949measurement,hansen1989relationships,hansen1990antenna} by antennas and RIS elements. Importantly, the number of parameters is hence independent of the RE's complexity and only depends on the number of attached ports $N_\mathrm{RE} = N_\mathrm{T} + N_\mathrm{R} + N_\mathrm{S}$, where $N_\mathrm{T}$, $N_\mathrm{R}$ and $N_\mathrm{S}$ denote the number of transmitting antennas, receiving antennas and RIS elements, respectively. Moreover, as long as $\mathbf{S}^\mathrm{RE}$ is known or can be estimated\footnote{See Ref.~\cite{sol2023experimentally,del2024minimal,V2NA_2p0} for various experimentally realized methods for physics-compliant end-to-end channel estimation, with varying degrees of parameter ambiguities. Irrespective of these ambiguities, the end-to-end wireless channels are correctly predicted as a function of D-RIS configuration in all cases. Further discussion is provided in Sec.~\ref{subsec_chanest}.}, no explicit description of the RE (in terms of geometry, materials, etc.) is necessary. These two important aspects first emerged in PhysFad-based works~\cite{prod2023efficient,sol2023experimentally} and were subsequently transposed to an equivalent multi-port network formulation in Ref.~\cite{tapie2023systematic} where all required model parameters were extracted from a single full-wave simulation for an overwhelmingly complicated D-RIS-parametrized on-chip RE featuring reflective walls, extended dielectric layers and strong structural scattering. 

The RE's $N_\mathrm{S}$ auxiliary ports associated with the $N_\mathrm{S}$ RIS elements are terminated by a tunable load circuit. Hence, the load circuit (denoted by L in the following) has $N_\mathrm{S}$ ports and is characterized by a scattering matrix $ \mathbf{S}^\mathrm{L}\in\mathbb{C}^{N_\mathrm{S}\times N_\mathrm{S}}$. In the case of a D-RIS, $\mathbf{S}^\mathrm{L}$ is diagonal and contains the reflection coefficients of the $N_\mathrm{S}$ tunable lumped elements. In the case of a BD-RIS, $\mathbf{S}^\mathrm{L}$ is ``beyond-diagonal'' and includes the scattering matrix associated with the reconfigurable impedance network.

Given $ \mathbf{S}^\mathrm{RE}$ and $\mathbf{S}^\mathrm{L}$, the standard ``cascade loading'' formula from multi-port network theory~\cite{anderson_cascade_1966,ha1981solid,ferrero1992new,prod2024efficient} (see Appendix~B.3 in Ref.~\cite{prod2024efficient} for a recent derivation) directly yields the resulting scattering matrix $\Tilde{\mathbf{S}} \in \mathbb{C}^{N_\mathrm{A}\times N_\mathrm{A}}$, where $N_\mathrm{A} = N_\mathrm{T}+N_\mathrm{R}$, that is ``measurable'' via the antenna ports in the RE:
\begin{equation}
    \Tilde{\mathbf{S}} = \mathbf{S}^\mathrm{RE}_\mathcal{AA} +\mathbf{S}^\mathrm{RE}_\mathcal{AS} \left(\left( \mathbf{S}^\mathrm{L}\right)^{-1} - \mathbf{S}^\mathrm{RE}_\mathcal{SS}  \right)^{-1} \mathbf{S}^\mathrm{RE}_\mathcal{SA},
    \label{eq_new1}
\end{equation}
where $\mathcal{A}$ and $\mathcal{S}$ denote the sets of port indices of the RE associated with the antennas and RIS elements, respectively.

The end-to-end wireless channel matrix $\mathbf{H}\in\mathbb{C}^{N_\mathrm{R} \times N_\mathrm{T}}$ is then simply an off-diagonal block of $\Tilde{\mathbf{S}}$:
\begin{equation}
    \mathbf{H} = \left[  \Tilde{\mathbf{S}} \right]_\mathcal{RT} = \mathbf{S}^\mathrm{RE}_\mathcal{RT} +\mathbf{S}^\mathrm{RE}_\mathcal{RS} \left(\left( \mathbf{S}^\mathrm{L}\right)^{-1} - \mathbf{S}^\mathrm{RE}_\mathcal{SS}  \right)^{-1} \mathbf{S}^\mathrm{RE}_\mathcal{ST},
    \label{eq2_new}
\end{equation}
where $\mathcal{R}$ and $\mathcal{T}$ denote the sets of port indices associated with receiving and transmitting antennas, respectively, and $\mathcal{A} = \mathcal{T} \cup \mathcal{R}$. Importantly, we remark that in this physics-compliant end-to-end channel model, the dependence of $\mathbf{H}$ on the configuration of the RIS (encoded in $\mathbf{S}^\mathrm{L}$) is non-linear due to the matrix inversion that captures mutual coupling -- irrespective of whether we consider a D-RIS or a BD-RIS.

To connect this physics-compliant model to the widespread simplified cascaded model, we rewrite the matrix inversion in Eq.~(\ref{eq2_new}) as infinite sum of matrix powers:
\begin{equation}
    \begin{split}
        \mathbf{H} &= \mathbf{S}^\mathrm{RE}_\mathcal{RT} +\mathbf{S}^\mathrm{RE}_\mathcal{RS} \left(\left( \mathbf{S}^\mathrm{L}\right)^{-1} - \mathbf{S}^\mathrm{RE}_\mathcal{SS}  \right)^{-1} \mathbf{S}^\mathrm{RE}_\mathcal{ST} \\&= \mathbf{S}^\mathrm{RE}_\mathcal{RT} +\mathbf{S}^\mathrm{RE}_\mathcal{RS} \left( \mathbf{I}_{N_\mathrm{S}} -\mathbf{S}^\mathrm{L}\mathbf{S}^\mathrm{RE}_\mathcal{SS}  \right)^{-1}\mathbf{S}^\mathrm{L} \mathbf{S}^\mathrm{RE}_\mathcal{ST}\\&= \mathbf{S}^\mathrm{RE}_\mathcal{T} +\mathbf{S}^\mathrm{RE}_\mathcal{RS} \left( \sum_{k=0}^\infty  \left(\mathbf{S}^\mathrm{L}\mathbf{S}^\mathrm{RE}_\mathcal{SS}\right)^k  \right)\mathbf{S}^\mathrm{L} \mathbf{S}^\mathrm{RE}_\mathcal{ST},
    \end{split}
    \label{eqnew3}
\end{equation}
where the invertibility of $\left( \mathbf{I}_{N_\mathrm{S}} -\mathbf{S}^\mathrm{L}\mathbf{S}^\mathrm{RE}_\mathcal{SS}  \right)$ and the convergence of the infinite series are guaranteed by two basic assumptions: First, RE is a passive sub-unitary system, i.e., it does not involve any gain but inevitably features some absorption and/or leakage,\footnote{Absorption is the irreversible conversion of wave energy to other forms of energy like heat; gain would be the opposite. Meanwhile, leakage refers to wave energy exiting the system without passing through the considered ports.} such that $\mathbf{S}^\mathrm{RE}_\mathcal{SS}$ is strictly sub-unitary; second, L is a passive system that may be considered lossless and hence unitary in some theoretical studies but would be sub-unitary in any practical realization due to inevitable absorption, such that $\mathbf{S}^\mathrm{L}_\mathcal{SS}$ is unitary or sub-unitary. Hence, due to submultiplicativity, the spectral radius of $\mathbf{S}^\mathrm{L}\mathbf{S}^\mathrm{RE}_\mathcal{SS}$ is always below unity and the series always converges. For the same reasons, $\mathbf{S}^\mathrm{L}\mathbf{S}^\mathrm{RE}_\mathcal{SS} $ can never have an eigenvalue of 1 such that $\left( \mathbf{I}_{N_\mathrm{S}} -\mathbf{S}^\mathrm{L}\mathbf{S}^\mathrm{RE}_\mathcal{SS}  \right)$ can never be a singular matrix. 

Upon truncation of the infinite series in the last line of Eq.~(\ref{eqnew3}) after its first term, i.e., $k=0$, we obtain 
\begin{equation}
    \mathbf{H} \approx \mathbf{S}^\mathrm{RE}_\mathcal{RT} +\mathbf{S}^\mathrm{RE}_\mathcal{RS}   \mathbf{S}^\mathrm{L} \mathbf{S}^\mathrm{RE}_\mathcal{ST},
    \label{eqnew4}
\end{equation}
which is the widespread simplified cascaded model of RIS-parametrized channels, decomposing the end-to-end channel into a contribution that did not encounter the RIS (the first term), and a contribution that interacted once with the RIS (the second term); $\mathbf{S}^\mathrm{L}$ is diagonal for D-RIS and ``beyond-diagonal'' for BD-RIS. Note that the truncation at $k=0$ is potentially a very strong assumption and Eq.~(\ref{eqnew4}) is generally \textit{not} physics-compliant.

The principle of this derivation of the widespread simplified cascaded model from the physics-compliant model was first proposed for D-RIS based on PhysFad in Ref.~\cite{rabault2023tacit}; here, we have transposed it to multi-port network theory using scattering parameters, and extended it to the BD-RIS case.

\subsection{Key Insight}
\label{sub_sec_keyinsight}

\begin{figure*}[!h]
    \centering
    \includegraphics[width=\textwidth]{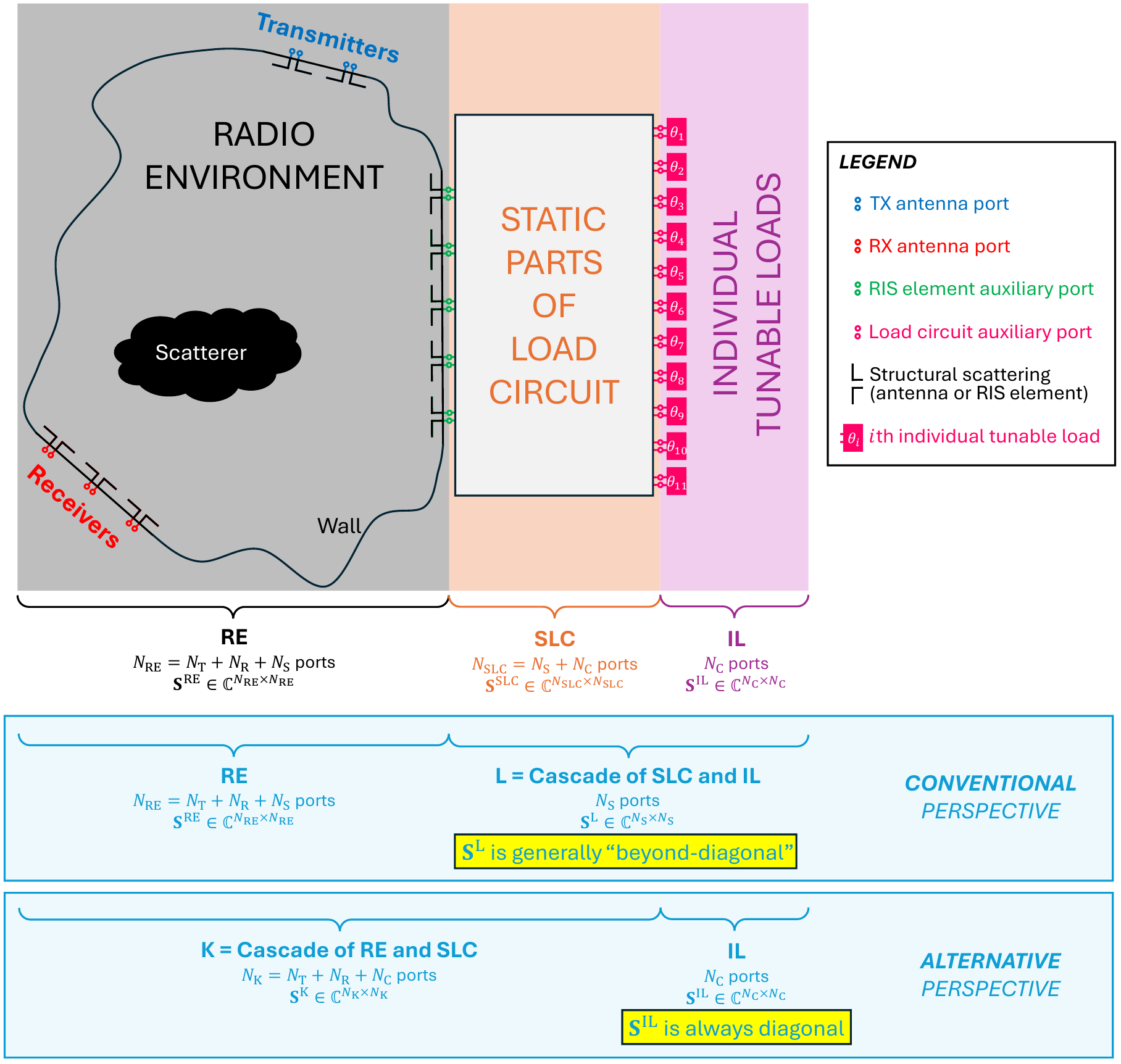}
    \caption{Generic multi-port chain cascade analysis of BD-RIS-parametrized wireless channels. The radio environment (RE) is an $N_\mathrm{RE}$-port system characterized by the scattering matrix $\mathbf{S}^\mathrm{RE}$, where $N_\mathrm{RE}$ is the sum of $N_\mathrm{T}$ (number of transmitting antennas), $N_\mathrm{R}$ (number of receiving antennas) and $N_\mathrm{S}$ (number of RIS elements). $\mathbf{S}^\mathrm{RE}$ lumps together all scattering effects in the RE, including scattering originating from objects and walls as well as the structural scattering by the antennas and RIS elements. The $N_\mathrm{S}$ auxiliary ports associated with the RIS elements are terminated by a load circuit (L). L can be separated into its static parts (SLC) and individual tunable lumped elements (IL), characterized by the scattering matrices $\mathbf{S}^\mathrm{SLC}$ and $\mathbf{S}^\mathrm{IL}$, respectively. SLC and IL have $N_\mathrm{SLC}=N_\mathrm{S}+N_\mathrm{C}$ and $N_\mathrm{C}$ ports, respectively. 
    The $N_\mathrm{C}$ ports of SLC associated with tunable lumped elements are terminated by IL.  (In the case of a conventional D-RIS, each RIS element auxiliary port is directly and uniquely connected to one individual tunable load, implying $N_\mathrm{C}=N_\mathrm{S}$ and $\mathbf{S}^\mathrm{SLC} = \left[ \mathbf{0}_{N_\mathrm{S}} \ \  \mathbf{I}_{N_\mathrm{S}};  \mathbf{I}_{N_\mathrm{S}}  \ \ \mathbf{0}_{N_\mathrm{S}}\right] $.) The conventional perspective on this chain cascade (albeit never explicited so far) consists in first evaluating the cascade of SLC and IL, referred to as L, which yields a \textit{generally ``beyond-diagonal''} scattering matrix $\mathbf{S}^\mathrm{L}$ that terminates the RIS element auxiliary ports of the RE. The alternative perspective emphasized in this paper is to first evaluate the cascade of RE and SLC, referred to as K; the auxiliary ports of K associated with tunable lumped elements are then terminated by $\mathbf{S}^\mathrm{IL}$ which is \textit{always diagonal}. }
    \label{Fig1}
\end{figure*}

Our key insight is that the load circuit (L) of a BD-RIS, irrespective of its architectural details, can always be separated into a static part (SLC) and a set of tunable individual lumped elements (IL). We denote the number of tunable elements within the BD-RIS load circuit by $N_\mathrm{C}$; typically, $N_\mathrm{C} > N_\mathrm{S}$ but our theory holds equally for $N_\mathrm{C} \leq N_\mathrm{S}$. It follows that LC must be amenable to a description as a multi-port network with $N_\mathrm{SLC} = N_\mathrm{S} + N_\mathrm{C}$ ports, of which $N_\mathrm{S}$ are connected to the RE's auxiliary ports associated with RIS elements and $N_\mathrm{C}$ are terminated by the individual tunable loads comprised in IL. Consequently, L is the result of cascade-loading SLC with IL, analogous to an RE parametrized by a D-RIS. By analogy with Eq.~(\ref{eq_new1}), it directly follows that
\begin{equation}
    \mathbf{S}^\mathrm{L} = \mathbf{S}^\mathrm{SLC}_{\bar{\mathcal{S}}\bar{\mathcal{S}}}+\mathbf{S}^\mathrm{SLC}_{\bar{\mathcal{S}}\mathcal{C}} \left(\left( \mathbf{S}^\mathrm{IL}\right)^{-1} - \mathbf{S}^\mathrm{SLC}_\mathcal{CC}  \right)^{-1} \mathbf{S}^\mathrm{SLC}_{\mathcal{C}\bar{\mathcal{S}}},
    \label{eq_new5}
\end{equation}
where $\mathbf{S}^\mathrm{SLC} \in \mathbb{C}^{N_\mathrm{SLC} \times N_\mathrm{SLC}}$ is the scattering matrix of SLC and $\mathbf{S}^\mathrm{IL} \in \mathbb{C}^{N_\mathrm{C} \times N_\mathrm{C}}$ is the scattering matrix of IL. $\mathbf{S}^\mathrm{IL}$ is diagonal; its $i$th entry is the reflection coefficient of the $i$th tunable lumped element within the load circuit. $\bar{\mathcal{S}}$ and $\mathcal{C}$ denote the sets of port indices of SLC associated with RIS elements and tunable individual loads, respectively.

At this stage, the following three remarks are important:
\begin{itemize}
    \item The physics-compliant decomposition of the BD-RIS load circuit into static parts and tunable lumped elements, as presented in Eq.~(\ref{eq_new5}), underpins the key insight of this paper. This decomposition was thus far overlooked in the BD-RIS literature.
    \item A D-RIS is a special case of this generic description, For a D-RIS, $N_\mathrm{S} = N_\mathrm{C}$ and SLC takes the trivial form of connecting each RIS element auxiliary port directly and uniquely to one tunable lumped element. Mathematically, in the case of a D-RIS, we thus have $\mathbf{S}^\mathrm{SLC} = \left[ \mathbf{0}_{N_\mathrm{S}} \ \  \mathbf{I}_{N_\mathrm{S}}; \ \mathbf{I}_{N_\mathrm{S}}  \ \ \mathbf{0}_{N_\mathrm{S}}\right] $.
    \item We dedicate Sec.~\ref{sec_determining_S_LC} to explaining how to determine $\mathbf{S}^\mathrm{SLC}$ in any given theoretical, numerical or experimental study involving BD-RIS. However, operationally, it is usually not necessary to explicitly determine $\mathbf{S}^\mathrm{SLC}$, as seen in our case study in Sec.~\ref{sec_e2e_estim}.
\end{itemize}

Given the insight that L is a cascade of SLC and IL, it follows that the end-to-end BD-RIS-parametrized wireless channel arises in fact from the chain cascade of three multi-port systems: RE, SLC and IL. A detailed illustration thereof is provided in Fig.~\ref{Fig1}. Conventionally, the separation of L into SLC and IL is not considered. In other words, the conventional representation of BD-RIS-parametrized channels implicitly assumes that SLC and IL have already been cascaded, yielding L, such that one is left with the cascade loading of RE by L. In this representation, one works directly with Eq.~(\ref{eq2_new}) in which $\mathbf{S}^\mathrm{L}$ is a ``beyond-diagonal'' matrix.

The alternative representation of BD-RIS-parametrized channels proposed in this paper starts instead by evaluating the cascade of RE and LC, yielding a system (named K) encapsulating all static parts within the chain cascade; then, the auxiliary ports of K associated with the tunable lumped elements are terminated by IL. K is characterized by
\begin{equation}
  \mathbf{S}^\mathrm{K} = \begin{bmatrix}  \mathbf{S}^\mathrm{K}_\mathcal{AA} &  \mathbf{S}^\mathrm{K}_\mathcal{AC}\\  \mathbf{S}^\mathrm{K}_\mathcal{CA} &  \mathbf{S}^\mathrm{K}_\mathcal{CC}   \end{bmatrix}  \in \mathbb{C}^{ N_\mathrm{K} \times N_\mathrm{K} },  
  \label{eq7}
\end{equation}
where $N_\mathrm{K} = N_\mathrm{A}+N_\mathrm{C}$. The analytical expression for $\mathbf{S}^\mathrm{K}$ is obtained via the well-established ``Redheffer star product'' \cite{redheffer_inequalities_1959,chu_generalized_1986,overfelt1989alternate,prod2024efficient}:
\begin{equation}
\begin{split}
& {\mathbf{S}}^{\mathrm{K}}_{\mathcal{A}\mathcal{A}} = \mathbf{S}^\mathrm{RE}_{\mathcal{A}\mathcal{A}} - \mathbf{S}^\mathrm{RE}_{\mathcal{A}\mathcal{S}} \mathbf{S}^\mathrm{LC}_{\bar{\mathcal{S}}\bar{\mathcal{S}}} \mathbf{X}_1 \mathbf{S}^\mathrm{RE}_{\mathcal{S}\mathcal{A}},\\
& {\mathbf{S}}^{\mathrm{K}}_{\mathcal{A}\mathcal{C}} = -\mathbf{S}^\mathrm{RE}_{\mathcal{A}\mathcal{S}} \mathbf{X}_2 \mathbf{S}^\mathrm{LC}_{\bar{\mathcal{S}}\mathcal{C}},\\
& {\mathbf{S}}^{\mathrm{K}}_{\mathcal{C}\mathcal{A}} = -\mathbf{S}^\mathrm{LC}_{\mathcal{C}\bar{\mathcal{S}}} \mathbf{X}_1 \mathbf{S}^\mathrm{RE}_{\mathcal{S}\mathcal{A}},\\
& {\mathbf{S}}^{\mathrm{K}}_{\mathcal{C}\mathcal{C}} = \mathbf{S}^\mathrm{LC}_{\mathcal{C}\mathcal{C}} - \mathbf{S}^\mathrm{LC}_{\mathcal{C}\bar{\mathcal{S}}} \mathbf{S}^\mathrm{RE}_{\mathcal{S}\mathcal{S}} \mathbf{X}_2 \mathbf{S}^\mathrm{LC}_{\bar{\mathcal{S}}\mathcal{C}},
\end{split}
  \label{eq8}
\end{equation}
where
\begin{equation}
\begin{split}
 & \mathbf{X}_1 = \left( \mathbf{S}^\mathrm{RE}_{\mathcal{S}\mathcal{S}} \mathbf{S}^\mathrm{LC}_{\bar{\mathcal{S}}\bar{\mathcal{S}}} - \mathbf{I}_{N_\mathrm{S}} \right)^{-1}, \\
& \, \mathbf{X}_2 = \left( \mathbf{S}^\mathrm{LC}_{\bar{\mathcal{S}}\bar{\mathcal{S}}} \mathbf{S}^\mathrm{RE}_{\mathcal{S}\mathcal{S}} - \mathbf{I}_{N_\mathrm{S}} \right)^{-1}.
\end{split}
  \label{eq6}
\end{equation}
This expression for $\mathbf{S}^{\mathrm{K}}$ is mathematically more complex than that for $\mathbf{S}^{\mathrm{L}}$ in Eq.~(\ref{eq_new5}) because both RE and LC have ``free'' (unconnected) ports after their connection whereas only LC but not IL has free ports after their connection~\cite{prod2024efficient}.

Given $ \mathbf{S}^\mathrm{K}$ and $\mathbf{S}^\mathrm{IL}$, the standard ``cascade loading'' formula directly yields $\Tilde{\mathbf{S}}$; specifically, analogous to Eq.~(\ref{eq_new1}), we obtain
\begin{equation}
    \tilde{\mathbf{S}} = \mathbf{S}^\mathrm{K}_\mathcal{AA} +\mathbf{S}^\mathrm{K}_\mathcal{AC} \left(\left( \mathbf{S}^\mathrm{IL}\right)^{-1} - \mathbf{S}^\mathrm{K}_\mathcal{CC}  \right)^{-1} \mathbf{S}^\mathrm{K}_\mathcal{CA}
    \label{eq10}
\end{equation}
and hence 
\begin{equation}
    \mathbf{H} = \left[  \Tilde{\mathbf{S}} \right]_\mathcal{RT} = \mathbf{S}^\mathrm{K}_\mathcal{RT} +\mathbf{S}^\mathrm{K}_\mathcal{RC} \left(\left( \mathbf{S}^\mathrm{IL}\right)^{-1} - \mathbf{S}^\mathrm{K}_\mathcal{CC}  \right)^{-1} \mathbf{S}^\mathrm{K}_\mathcal{CT}.
    \label{eq22_new}
\end{equation}
Importantly, $\mathbf{S}^\mathrm{IL}$ is a diagonal matrix such that the mathematical structure of Eq.~(\ref{eq22_new}) for BD-RIS is analogous to that of Eq.~(\ref{eq_new1}) in the D-RIS case. This is our key result.

Obviously, the expressions for $\mathbf{H}$ from Eq.~(\ref{eq22_new}) must be equal to that from Eq.~(\ref{eq2_new}) since we did not make any assumptions to obtain either of these equations. It is instructive to compare the role of corresponding quantities in these equations:
\begin{subequations}
\begin{equation}
        \mathrm{Eq.(\ref{eq2_new})}  \leftrightarrow \mathrm{Eq.(\ref{eq22_new})},
\end{equation}\begin{equation}
        \mathbf{S}^\mathrm{RE}  \leftrightarrow \mathbf{S}^\mathrm{K},
\end{equation}
\begin{equation}
            \mathbf{S}^\mathrm{L}  \leftrightarrow \mathbf{S}^\mathrm{IL},
\end{equation}
\begin{equation}
            \mathcal{S}  \leftrightarrow \mathcal{C}.
\end{equation}
\label{eq_analogy_mpn_bdris}
\end{subequations}
In the conventional representation, the primary ports are those comprised in $\mathcal{A}\cup\mathcal{S}$, and the ports of RE from $\mathcal{S}$ are terminated by  $\mathbf{S}^\mathrm{L}$ which is \textit{generally ``beyond-diagonal''}. In our proposed alternative representation, the primary ports are those comprised in $\mathcal{A}\cup\mathcal{C}$, and the ports of K within $\mathcal{C}$ are terminated by $\mathbf{S}^\mathrm{IL}$ which is \textit{always diagonal}.
We reiterate that the conventional representation and our alternative representation are both physics-compliant and fully equivalent. However, the alternative representation allows us to directly apply physics-compliant D-RIS algorithms to BD-RIS scenarios, as we demonstrate in Sec.~\ref{sec_e2e_estim}.

\subsection{Transposition to Coupled-Dipole Model (PhysFad) for RIS-Parametrized Channels}
\label{sec_physfad}

Thus far, we have only discussed circuit-theory-based physics-compliant modeling of BD-RIS parametrized channels in terms of scattering parameters. For the D-RIS case, an alternative physics-compliant approach based on a coupled-dipole formalism exists. It was introduced as PhysFad~\cite{faqiri2022physfad} and constituted the basis of the first experimental validation of a physics-compliant channel model for D-RIS-parametrized rich-scattering channels as well as various frugal physics-compliant end-to-end channel estimation protocols~\cite{sol2022meta}. In this subsection, we first review the basics of PhysFad (Sec.~\ref{subsec_physfad_background}). Then, we establish the operational equivalence between PhysFad and scattering-parameter-based circuit models for D-RIS (Sec.~\ref{subsec_OpEq}). Finally, combining Sec.~\ref{subsec_OpEq} with our diagonal scattering-parameter-based representation of BD-RIS-parametrized channels from Sec.~\ref{sub_sec_keyinsight}, we will describe BD-RIS-parametrized channels within the PhysFad formalism (Sec.~\ref{subsubsec_physfad_bdris}).

\subsubsection{Background on PhysFad for D-RIS}
\label{subsec_physfad_background}

PhysFad assigns a dipole to each of the $N=N_\mathrm{A}+N_\mathrm{S}$ ports of the primary wireless entities: $N_\mathrm{A}$ antenna ports and $N_\mathrm{S}$ auxiliary ports associated with each RIS elements' tunable lumped elements. The $i$th dipole is characterized by its polarizability $\alpha_i$, which is tunable in the case of the tunable lumped elements. In the compact formulation of PhysFad~\cite{prod2023efficient,sol2023experimentally,sol2024optimal}, the $i$th and $j$th dipoles are coupled to each other via the \textit{background} Green's function $G_{ij}=G_{ji}$ which takes into account all scattering effects within the radio environment (scattering objects of any size and material, including structural scattering of antennas and RIS elements).

Then, an interaction matrix $\mathbf{W} \in \mathbb{C}^{N \times N}$ is set up, composed of the sum of a diagonal matrix containing the inverse polarizabilities and a symmetric matrix containing the background Green's functions:
\begin{equation}
W_{i,j}=\begin{cases}
\alpha_{i}^{-1}, & i=j\\
-G_{ij}, & i\neq j
\end{cases}.
\end{equation}
In PhysFad, we define the RIS configuration as the difference between some choice of reference inverse polarizability $\alpha_0^{-1}$ and the RIS elements' inverse polarizabilities:
\begin{equation}
   \mathbf{c} =  \left[ \alpha_0^{-1} - \alpha_i^{-1} \  \vert \  i \in \mathcal{S} \right].
\end{equation}
A convenient partioned representation of $\mathbf{W}(\mathbf{c})$ follows:
\begin{equation}
    \mathbf{W}(\mathbf{c})= \begin{bmatrix} 
	\mathbf{W}^0_{\mathcal{AA}}  & \mathbf{W}^0_{\mathcal{AS}} \\
	\mathbf{W}^0_{\mathcal{SA}}  & \mathbf{W}^0_{\mathcal{SS}}-\mathrm{diag}(\mathbf{c}) \\
\end{bmatrix}.
\end{equation}
The end-to-end channel matrix is finally identified as the $\mathcal{RT}$ block of the inverse of this interaction matrix~\cite{faqiri2022physfad}:
\begin{equation}
    \mathbf{H}_\mathrm{PF}(\mathbf{c}) = \left[ \mathbf{W}^{-1}(\mathbf{c}) \right]_\mathcal{RT}.
    \label{eq20}
\end{equation}

Whereas circuit theory treats systems as black boxes, PhysFad's coupled-dipole approach describes the physical wave-matter interactions inside the systems, enabling insights that are not accessible via circuit theory such as the method for performing conjugate beamforming without pilot exchange purely based on user motion that was proposed and experimentally demonstrated in Ref.~\cite{sol2024optimal}.

\subsubsection{Operational Equivalence}
\label{subsec_OpEq}

We now elaborate on the operational equivalence between the scattering-parameter-based circuit representation used in Sec.~\ref{subsec_background} and the PhysFad representation described in Sec.~\ref{subsec_physfad_background}. By ``operational equivalence'', we mean that PhysFad has the same mathematical structure as the multi-port network formulation, such that end-to-end channel estimation and optimization for D-RIS are equivalent with both formulations. 

We define $\mathbf{\Omega}^0 = \mathbf{W}^{-1}(\mathbf{c}=\mathbf{0})$ and, after a few basic linear algebra steps detailed in Appendix~\ref{AppendixB}, we obtain
\begin{equation}
  \mathbf{H}_\mathrm{PF}(\mathbf{c}) =   \mathbf{\Omega}^0_\mathcal{RT} + \mathbf{\Omega}^0_\mathcal{RS} \left( \mathbf{\Phi} - \mathbf{\Omega}^0_\mathcal{SS} \right)^{-1} \mathbf{\Omega}^0_\mathcal{ST}.
  \label{eq_21}
\end{equation}
The mathematical structure of Eq.~(\ref{eq_21}) based on PhysFad is identical to that of Eq.~(\ref{eq2_new}) based on multi-port network theory, and the following correspondences are apparent:
\begin{subequations}
\begin{equation}
        \mathrm{Eq.~(\ref{eq2_new})}  \leftrightarrow \mathrm{Eq.~(\ref{eq_21})},
\end{equation}\begin{equation}
        \mathbf{S}^\mathrm{RE}  \leftrightarrow \mathbf{\Omega}^0,
\end{equation}
\begin{equation}
            \mathbf{S}^\mathrm{L}  \leftrightarrow \mathbf{\Phi}.
\end{equation}
\label{eq_analogy_mpn_physfad}
\end{subequations}

\subsubsection{PhysFad for BD-RIS}
\label{subsubsec_physfad_bdris}

By comparing Eq.~(\ref{eq_analogy_mpn_bdris}) and Eq.~(\ref{eq_analogy_mpn_physfad}), it is immediately clear that our proposed diagonal representation for BD-RIS-parametrized channels from Sec.~\ref{sub_sec_keyinsight} allows us also to model BD-RIS within PhysFad. The physical interpretation is as follows: The dipoles in the compact formulation of PhysFad are associated with the ports of the primary wireless entities. In a BD-RIS-parametrized channel, these are the $N_\mathrm{A}$ antenna ports and the $N_\mathrm{C}$ auxiliary ports associated with the tunable individual lumped elements within the load circuit. Hence, the representation of a BD-RIS-parametrized radio environment in PhysFad involves $N_\mathrm{K}$ dipoles, of which the $N_\mathrm{C}$ dipoles associated with tunable lumped elements in the load circuit have tunable polarizabilities. The background Green's functions between these $N_\mathrm{K}$ dipoles account for the potentially highly complex wave scattering within K (cascade of RE and SLC).

The rest of this paper uses our diagonal representation in terms of scattering parameters within the circuit-theoretic model described in Sec.~\ref{sub_sec_keyinsight}. However, because of the direct mathematically equivalent structure evidenced in Eq.~(\ref{eq_analogy_mpn_physfad}), it is obvious that all details of our experimentally grounded case study on end-to-end channel estimation and optimization in Sec.~\ref{sec_e2e_estim} would be exactly identical if the PhysFad formalism from this subsection was used instead.

\section{Determining $\mathbf{S}^\mathrm{SLC}$}
\label{sec_determining_S_LC}

At the core of our theoretical developments in Sec.~\ref{sub_sec_keyinsight} is $\mathbf{S}^\mathrm{SLC}$, the scattering matrix characterizing the static parts of the load circuit which connect the RIS elements to the tunable individual loads. The notion of $\mathbf{S}^\mathrm{SLC}$ has to date not been brought up explicitly in the BD-RIS literature, to the best of our knowledge, such that we dedicate this section to discussing how to determine $\mathbf{S}^\mathrm{SLC}$ in various scenarios that can arise in theoretical, numerical or experimental studies of BD-RIS. At the end of this section, we go on to argue that, in fact, operationally there is no need to explicitly know $\mathbf{S}^\mathrm{SLC}$; only an estimate of $\mathbf{S}^\mathrm{K}$ is needed but not its breakdown into $\mathbf{S}^\mathrm{RE}$ and $\mathbf{S}^\mathrm{SLC}$, as evidenced in our case study in Sec.~\ref{sec_e2e_estim}.

\subsection{Theoretical canonical examples}
\label{subsec_canonical_examples}

Theoretically studied tunable circuits that terminate RIS elements in BD-RIS typically include tunable lumped elements that are connected without loss, delay and dispersion to each other and/or the RIS element auxiliary ports and/or ground. Under these simplifying assumptions, $\mathbf{S}^\mathrm{SLC}$ can be determined analytically.

Thus far, existing theoretical studies of BD-RIS directly obtained $\mathbf{S}^\mathrm{L}$ without explicitly identifying $\mathbf{S}^\mathrm{SLC}$ and $\mathbf{S}^\mathrm{IL}$. More specifically, as detailed in Eq.~(13) of Ref.~\cite{shen2021modeling} (based on Eq.~(6) of Ref.~\cite{murch2016}), the admittance representation $\mathbf{Y}^\mathrm{L}$ of the circuit is obtained in closed form, which can then be transformed to $\mathbf{S}^\mathrm{L}$, if desired.

\begin{figure}[h]
    \centering
    \includegraphics[width=\linewidth]{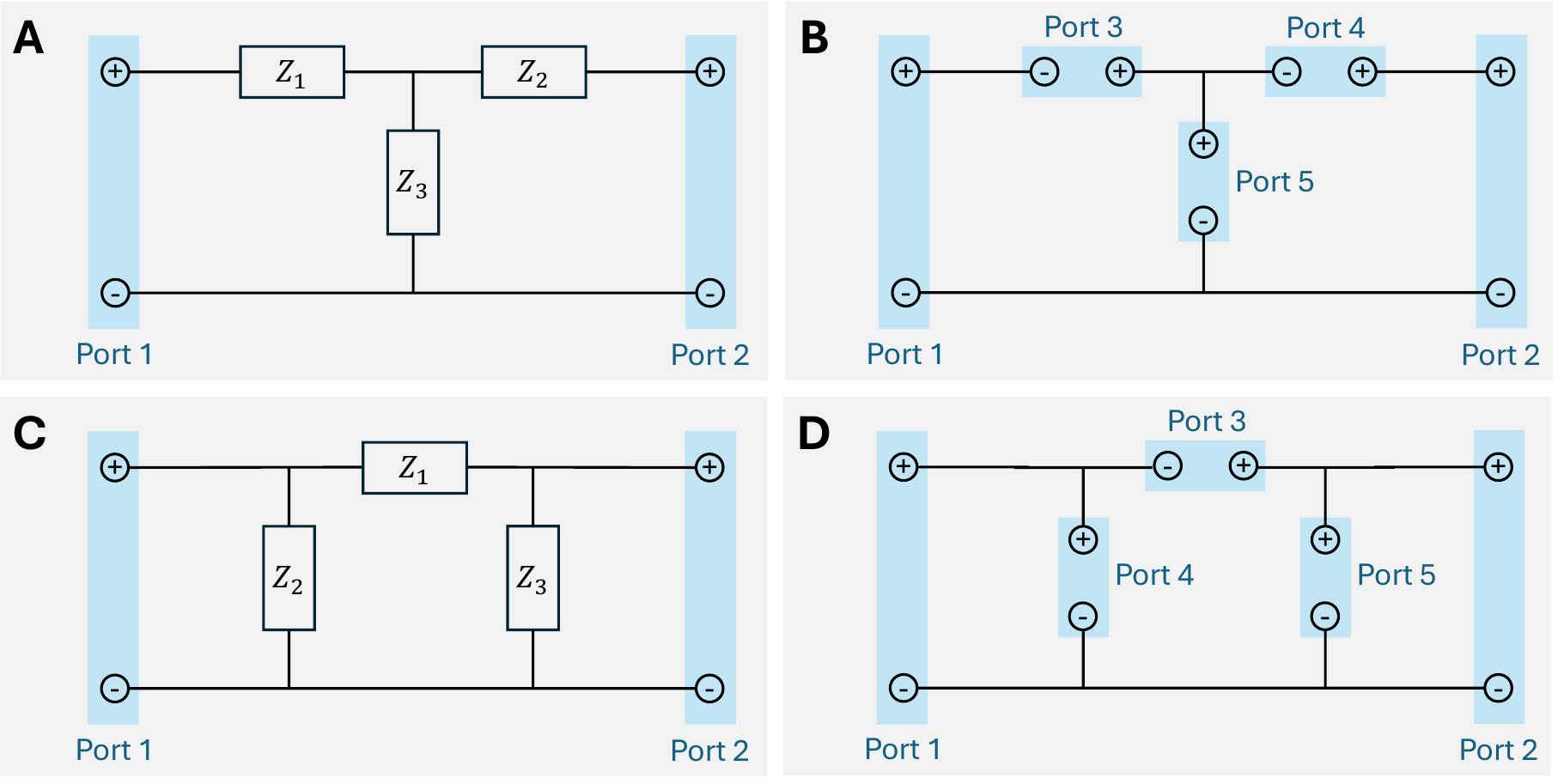}
    \caption{(A) Ideal 2-port T network. (B) The three impedances in (A) are replaced by auxiliary ports, yielding a 5-port network whose scattering matrix can be determined analytically, see Appendix~\ref{AppendixA} and Eq.~(\ref{eq_T}). (C) Ideal 2-port $\pi$ network. (D) The three impedances in (C) are replaced by auxiliary ports, yielding a 5-port network whose scattering matrix can be determined analytically, see Eq.~(\ref{eq_pi}). }
    \label{Fig2}
\end{figure}

We begin by analytically considering a simple ideal two-port T network containing three lumped impedances $Z_1$, $Z_2$ and $Z_3$, as sketched in Fig.~\ref{Fig2}A. The characteristic impedance of the two transmission lines attached to the two ports is $Z_0 = 50 \ \Omega$.  Recall that black lines denote connections without loss, delay or dispersion. 

Using basic circuit analysis, as detailed in Appendix~\ref{AppendixA}, we readily obtain the $5 \times 5$ scattering matrix of the circuit from Fig.~\ref{Fig2}B:
\begin{equation}
    \mathbf{S}^\mathrm{SLC,T} = \begin{bmatrix}
0.25 & 0.25 & -0.75 & -0.25 & 0.5 \\
0.25 & 0.25 & 0.25 & 0.75 & 0.5 \\
-0.75 & 0.25 & 0.25 & -0.25 & 0.5 \\
-0.25 & 0.75 & -0.25 & 0.25 & -0.5 \\
0.5 & 0.5 & 0.5 & -0.5 & 0
\end{bmatrix}.
\label{eq_T}
\end{equation}
We have verified this result with three sanity checks, as detailed in Appendix~\ref{AppendixA}.

A similar analysis for a $\pi$ network whose three lumped impedances have been replaced by auxiliary ports yields
\begin{equation}
    \mathbf{S}^\mathrm{SLC,\pi} = \begin{bmatrix}
-0.25 & 0.25 & -0.5 & 0.75 & 0.25 \\
0.25 & -0.25 & 0.5 & 0.25 & 0.75 \\
-0.5 & 0.5 & 0 & -0.5 & 0.5 \\
0.75 & 0.25 & -0.5 & -0.25 & 0.25 \\
0.25 & 0.75 & 0.5 & 0.25 & -0.25
\end{bmatrix}.
\label{eq_pi}
\end{equation}
Again, we verified this result with three sanity checks. A reconfigurable ideal $\pi$ network like the one we study was considered in Fig.~3(a) of Ref.~\cite{shen2021modeling} for a group-connected BD-RIS with a group size of two.

The same exercise can be repeated for more complex ideal reconfigurable impedance networks such as the one from Fig.~2(b) in Ref.~\cite{shen2021modeling} but manual execution of the required circuit analysis rapidly becomes tiresome; already for the example from Fig.~2(b) in Ref.~\cite{shen2021modeling} the dimensions of the sought-after $\mathbf{S}^\mathrm{SLC}$ would be $14 \times 14$. Therefore, for studies of large and/or realistic (as opposed to ideal) reconfigurable impedance networks, the numerical or experimental approaches described in the next two subsections are more suitable.

\subsection{Numerical full-wave simulations}
\label{subsec_numerical}

Whereas the previous subsection covered theoretical ideal load circuits that can be analyzed analytically, we now turn our attention to load circuits that can be realized in practice. These can generally not avoid the effects of loss, delay and dispersion, and in most cases, an analytical expression for $\mathbf{S}^\mathrm{SLC}$ will not exist. Nonetheless, it is reasonable to assume that for practical studies of realistic BD-RIS designs a detailed layout of the circuit terminating the RIS elements is available. Indeed, such a layout will be an inevitable pre-requisite for any experimentally realized prototype fabrication. Consequently, this circuit can be simulated numerically in full-wave solvers. In fact, a single full-wave simulation is sufficient to obtain $\mathbf{S}^\mathrm{SLC}$ if one simply places lumped ports at the locations of the tunable lumped elements. Incidentally, this is exactly the procedure that was already implemented in Ref.~\cite{tapie2023systematic} to extract the analytically intractable scattering matrix of a highly complex on-chip RE parametrized by a D-RIS. As mentioned earlier, an RE parametrized by a D-RIS is conceptually exactly the same problem as a load circuit parametrized by tunable individual loads. To summarize, for any arbitrarily complex load circuit design, a single full-wave simulation can yield $\mathbf{S}^\mathrm{SLC}$.

\subsection{Experimental estimation }

In this subsection, we discuss the extent to which $\mathbf{S}^\mathrm{SLC}$ could be estimated experimentally. A pivotal requirement for such an experimental estimation would be the possibility to input and output waves through the load circuit ports connected to the RIS elements (i.e., ports of SLC whose indices are in the set $\bar{\mathcal{S}}$); this could be possible, for instance, before connecting the RIS and the load circuit. Then, the considered problem is analogous to the problem of physics-compliant end-to-end channel estimation with a D-RIS that was tackled in Refs.~\cite{sol2023experimentally,del2024minimal}. Indeed, the load circuit is a multi-port network characterized by $\mathbf{S}^\mathrm{SLC}$ that is terminated by individual tunable loads. Based on experimental measurements at the ports of the load circuit from set $\bar{\mathcal{S}}$ for different terminations of the ports from set $\mathcal{C}$, it is provably impossible to unambiguously determine $\mathbf{S}^\mathrm{SLC}$ experimentally because at least sign ambiguities on off-diagonal terms associated with the ports from set $\mathcal{C}$ will remain~\cite{del2024minimal,V2NA_2p0}. If not at least three distinct and characterized individual load states are available at the ports from $\mathcal{C}$, additional ambiguities about $\mathbf{S}^\mathrm{SLC}$ ensue. However, operationally, these ambiguities are not problematic because as long as $\mathbf{S}^\mathrm{L}$ is correctly predicted we have everything we need to evaluate $\mathbf{H}$ based on Eq.~(\ref{eq2_new}) by cascade-loading $\mathbf{S}^\mathrm{RE}$ at its ports comprised in  $\mathcal{S}$ with $\mathbf{S}^\mathrm{L}$. Indeed, Ref.~\cite{sol2023experimentally} accurately predicted wireless end-to-end channels in an unknown complex medium as a function of the binary configuration of a D-RIS prototype with unknown characteristics. The infinite number of satisfactory sets of parameter values should even facilitates the estimation of a plausible $\mathbf{S}^\mathrm{SLC}$ (and $\mathbf{S}^\mathrm{IL}$ if it is unknown) that yields an accurate prediction for $\mathbf{S}^\mathrm{L}$.

\subsection{Do we need to know $\mathbf{S}^\mathrm{SLC}$ explicitly?}

This section was dedicated to discussing various routes to determining $\mathbf{S}^\mathrm{SLC}$, a quantity that is at the core of the physics-compliant diagonal representation of BD-RIS parametrized channels put forth in the present paper. Although this section is pedagogically important to substantiate the theory developed in the previous section, arguably, \textit{in practice it is not even necessary to explicitly know $\mathbf{S}^\mathrm{SLC}$}. 

Indeed, in practice, the most compact modeling approach is our diagonal representation from Eq.~(\ref{eq22_new}) which directly maps the configuration of the individual tunable loads to the end-to-end wireless channels without separating K into RE and SLC. As long as we know $\mathbf{S}^\mathrm{K}$, operationally there is no need to break it down into $\mathbf{S}^\mathrm{RE}$ and $\mathbf{S}^\mathrm{SLC}$. The primary ports in a BD-RIS chain cascade (see Fig.~\ref{Fig1}) are those associated with the antennas and the tunable lumped elements. Meanwhile, the RIS elements (and the associated ports) are merely a detail of how these primary entities are coupled to each other. Therefore, in the next section, we propose a scheme for end-to-end channel estimation and optimization that only estimates $\mathbf{S}^\mathrm{K}$ and $\mathbf{S}^\mathrm{IL}$ in order to successfully predict $\mathbf{H}$ as a function of the BD-RIS configuration.

\section{Directly applying D-RIS channel estimation and optimization algorithms to BD-RIS}
\label{sec_e2e_estim}

The most significant implication of our diagonal representation of BD-RIS parametrized channels for signal processing and wireless communications is that it implies that physics-compliant D-RIS algorithms can be directly applied in BD-RIS scenarios. This section is dedicated to demonstrating the feasibility of directly applying physics-compliant end-to-end channel estimation and optimization algorithms originally developed for D-RIS to BD-RIS-parametrized channels, based on an experimentally grounded case study. We first describe the considered scenario, then channel estimation, and finally channel optimization.

\subsection{Considered scenario grounded in an experiment}
\label{subsec_exp}

\begin{figure}[b]
    \centering
    \includegraphics[width=\linewidth]{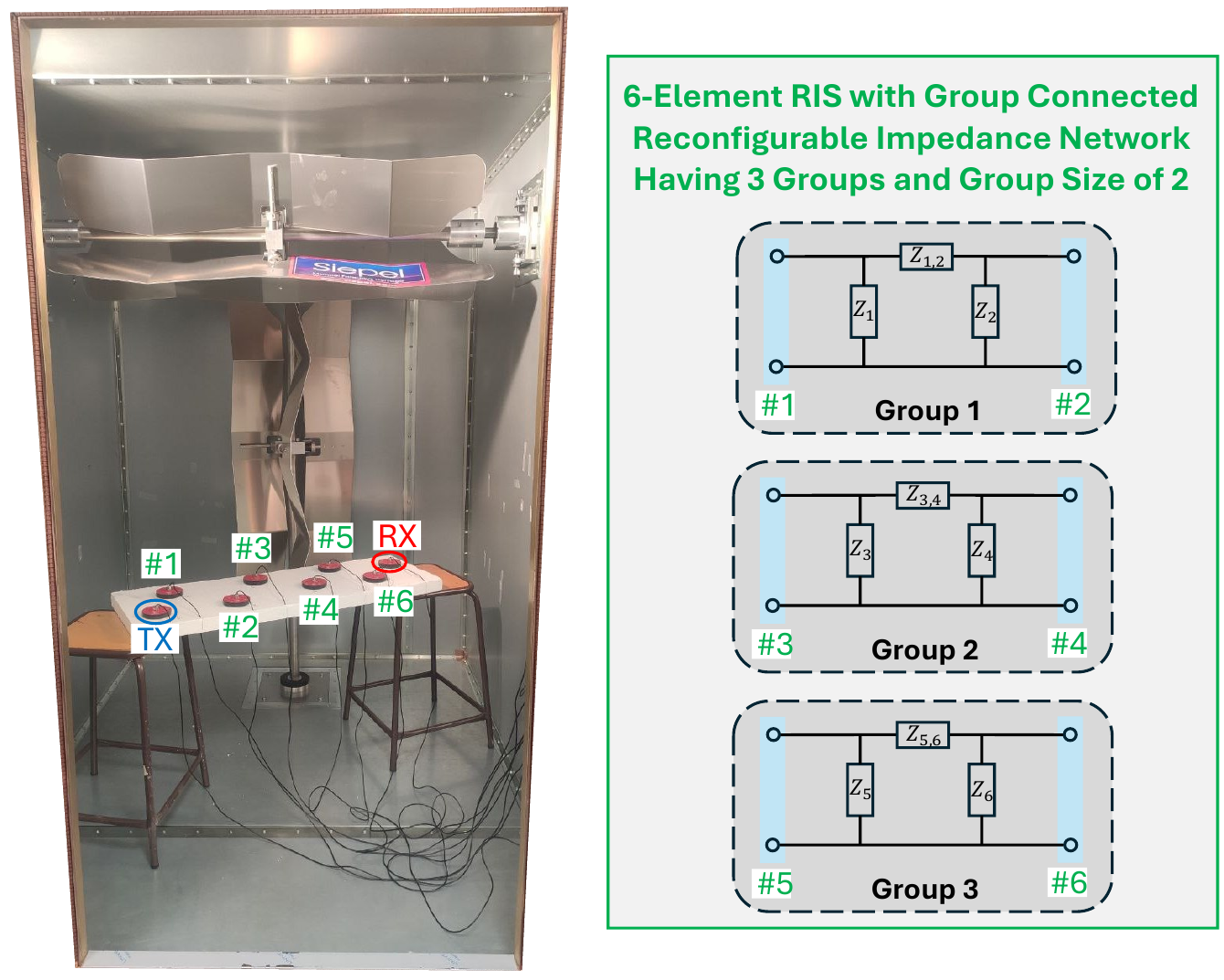}
    \caption{Experimentally grounded study of a SISO link inside a rich-scattering environment parametrized by a 6-element RIS with group-connected reconfigurable impedance network having three groups and group size two. }
    \label{Fig3}
\end{figure}

To ground the considered scenario in an experimental reality involving unknown rich scattering, we have placed eight commercial antennas (AEACBK081014-S698, designed for operation between 698~MHz and 2.69~GHz for GSM, SPRS and 4G (LTE)) inside a reverberation chamber (1.75~m~$\times$~1.50~m~$\times$~2.00~m) and measured the corresponding $8 \times 8$ scattering matrix with an eight-port vector network analyzer (Keysight M9005A) at $f=800$~MHz. The obtained scattering matrix is used as $\mathbf{S}^\mathrm{RE}$ in this case study. We consider a SISO link involving one transmitting and one receiving antenna, and the remaining six antennas are the RIS elements. The annotations on the photographic image in Fig.~\ref{Fig3} highlight this setup. As sketched in the right part of Fig.~\ref{Fig3}, the six RIS elements are split into three groups of size two that are considered to be connected to ideal $\pi$ networks, each of which is parametrized by three impedances. We assume that each impedance represents a PIN diode with two possible states whose characteristics are chosen according to those documented for a commercially available PIN diode (MADP-000907-14020x~\cite{macom}). Specifically, the two possible load impedances are $\eta^\mathrm{ON} = 5.2~\Omega$ and $\eta^\mathrm{OFF} = (\jmath\omega C)^{-1} = -7.96\jmath\times10^3~\Omega$, where $\jmath=\sqrt{-1}$, $\omega = 2\pi f$ and $C=25 \ \mathrm{fF}$. The corresponding reflection coefficients are $r_\mathrm{A} = (\eta^\mathrm{ON}-Z_0)/(\eta^\mathrm{ON}+Z_0) = -0.81$ and $r_\mathrm{B} = (\eta^\mathrm{OFF}-Z_0)/(\eta^\mathrm{OFF}+Z_0) = 0.9999 - 0.0126\jmath$. To recapitulate key parameters: $N_\mathrm{T}=1$, $N_\mathrm{R}=1$, $N_\mathrm{A}=N_\mathrm{T}+N_\mathrm{R}=2$, $N_\mathrm{S}=6$ and $N_\mathrm{C}=9$.

Our experimental measurements hence fix the entries of $\mathbf{S}^\mathrm{RE}$, Eq.~(\ref{eq_pi}) fixes the entries of the three diagonal blocks of the block-diagonal matrix $\mathbf{S}^\mathrm{SLC}$, and the PIN diode's data sheet fixes the two possible values that the diagonal entries of the diagonal matrix $\mathbf{S}^\mathrm{IL}$ can take. Specifically, for a given binary configuration $\mathbf{b}\in\mathbb{B}^{N_\mathrm{C}\times 1}$ of the $N_\mathrm{C}=9$ tunable load impedances, we can hence identify the corresponding $\mathbf{S}^\mathrm{IL}$ via
\begin{equation}
    \mathbf{S}^\mathrm{IL}(\mathbf{b}) = \mathrm{diag}\left( r_\mathrm{A} \mathbf{1}_{N_\mathrm{C}}+ (r_\mathrm{B}-r_\mathrm{A})\mathbf{b} \right).
    \label{eq_b}
\end{equation}
We can evaluate $\mathbf{S}^\mathrm{L}(\mathbf{b})$  by inserting $\mathbf{S}^\mathrm{IL}(\mathbf{b})$ (obtained with Eq.~(\ref{eq_b})) into Eq.~(\ref{eq_new5}) or via the known admittance matrix of an ideal $\pi$ network. Moreover, we can evaluate $\mathbf{S}^\mathrm{K}$ using Eq.~(\ref{eq8}). Finally, we can determine $\mathbf{H}(\mathbf{b}) $ via the conventional route using Eq.~(\ref{eq2_new}) based on $\mathbf{S}^\mathrm{RE}$ and $\mathbf{S}^\mathrm{L}(\mathbf{b}) $ or via our alternative route using Eq.~(\ref{eq22_new}) based on $\mathbf{S}^\mathrm{K}$ and $\mathbf{S}^\mathrm{IL}(\mathbf{b})$. We verified that both routes to evaluating $\mathbf{H}(\mathbf{b})$ yield exactly the same result for multiple random choices of $\mathbf{b}$. For the considered SISO problem, $\mathbf{H}(\mathbf{b}) $ is a scalar such that we denote it by $h(\mathbf{b}) $ in the following. 
Altogether, to the best of our knowledge, this constitutes the first study of a BD-RIS-parametrized wireless channel for which the utilized parameters are grounded in an experimental reality.

To perform physics-compliant end-to-end channel estimation and optimization, we simulate a scenario in which we can configure the reconfigurable impedance network to any desired binary configuration $\mathbf{b}$ and measure the corresponding wireless channel $h(\mathbf{b})$. Of course, we obtain $h(\mathbf{b})$ as detailed in the previous paragraph, but once  the pairs $\{\mathbf{b},h(\mathbf{b})\}$ are generated, we assume that we only know the pairs $\{\mathbf{b},h(\mathbf{b})\}$ and do not have any knowledge of $\mathbf{S}^\mathrm{RE}$, $\mathbf{S}^\mathrm{SLC}$, $r_\mathrm{A}$, or $r_\mathrm{B}$. We thereby simulate a true end-to-end channel estimation and optimization problem without any a priori knowledge of the characteristics of the radio environment, the load circuit or the tunable lumped elements. 

\subsection{End-to-end channel estimation}
\label{subsec_chanest}

The literature contains a few theoretical studies on channel estimation in scenarios involving BD-RIS~\cite{li2024channel,de2024channel}. However, these are based on simplified cascaded channel models. Here, we tackle the problem of physics-compliant end-to-end channel estimation of BD-RIS parametrized channels in a scenario that is grounded in experimental data. We highlight that thanks to the diagonal representation of the BD-RIS-parametrized channel we only need to estimate some version of $\mathbf{S}^\mathrm{K}$ and $\mathbf{S}^\mathrm{IL}$ but not the breakdown of $\mathbf{S}^\mathrm{K}$ into $\mathbf{S}^\mathrm{RE}$ and $\mathbf{S}^\mathrm{SLC}$. (The words ``some version'' in the previous sentence hint at the fact that ambiguities are inevitable and not problematic, as discussed below.)

\subsubsection{Problem statement}

Our ultimate goal in end-to-end channel estimation is to be able to map any conceivable $\mathbf{b}$ to the corresponding end-to-end wireless channel  $h(\mathbf{b})$:
\begin{equation}
    \mathbf{b} \rightarrow h(\mathbf{b}).
\end{equation}
As stated earlier, we do not assume to know anything but pairs of $\{\mathbf{b},h(\mathbf{b})\}$.

We hence face a parameter estimation problem. The most compact approach is based on the diagonal representation of the BD-RIS-parametrized channel in Eq.~(\ref{eq22_new}). It consists in determining the parameters necessary to predict $h$ using Eq.~(\ref{eq22_new}) for any given $\mathbf{b}$. These parameters are the blocks $\mathcal{RT}$, $\mathcal{AC}$ and $\mathcal{CC}$ of $\mathbf{S}^\mathrm{K}$ as well as the two possible values $r_\mathrm{A}$ and $r_\mathrm{B}$ that each diagonal entry of $\mathbf{S}^\mathrm{IL}$ can. 

Fortunately, this problem directly maps into the problem of physics-compliant end-to-end channel estimation for D-RIS for which experimentally validated solutions were already presented in Refs.~\cite{sol2022meta,del2024minimal}. Key insights from Refs.~\cite{sol2022meta,del2024minimal} as transposed to our BD-RIS-based scenario are summarized as follows:
\begin{enumerate}
    \item The number of parameters to be estimated is 
    \begin{equation}
    \begin{split}
        N_\mathrm{params} &= 2\left(2+\left[\frac{(N_\mathrm{A}+N_\mathrm{C})(N_\mathrm{A}+N_\mathrm{C}+1)}{2} - 2\right] \right) \\ &= 4 +\left[(N_\mathrm{A}+N_\mathrm{C})(N_\mathrm{A}+N_\mathrm{C}+1) - 4 \right]\\ &=132,
        \end{split}
    \end{equation}
    where the first term inside the brackets corresponds to the two possible reflection coefficients $r_\mathrm{A}$ and $r_\mathrm{B}$ of the individually tunable lumped elements, the second term inside the brackets corresponds to the number of unique entries in the blocks $\mathcal{RT}$, $\mathcal{AC}$ and $\mathcal{CC}$ of $\mathbf{S}^\mathrm{K}$, and the factor two in front of the brackets takes into account that the variables are complex-valued. Note that the number of parameters to be estimated neither depends on $N_\mathrm{S}$ nor on the complexity of the radio environment nor on the complexity of the load circuit.
    \item Any set of parameters that correctly predicts $h(\mathbf{b})$ is satisfactory. There are inevitable ambiguities in the parameter values because the parameter estimation problem is not sufficiently constrained to yield a unique solution. However, these ambiguities do not pose any problem with respect to our goal of accurately predicting $h(\mathbf{b})$.
    \item The parameter estimation problem can be solved in closed form~\cite{del2024minimal} or via gradient descent~\cite{sol2023experimentally,del2024minimal}. In the case of a gradient-descent approach, the parameters can be estimated purely based on non-coherent measurements~\cite{sol2023experimentally,del2024minimal}. Further benefits of the gradient-descent approach include its compatibility with opportunistic configuration switching and a better robustness against noise. The latter originates from the fact that the changes of $h$ between any two measurements are larger because on average half the tunable lumped elements have changed their state, as well as from the fact that any arbitrary number of measurements can readily be taken into account. 
\end{enumerate}

Here, we follow the gradient-descent approach put forth for the D-RIS scenario in Ref.~\cite{sol2023experimentally}.

\subsubsection{Algorithm}

Our physics-compliant end-to-end channel estimation algorithm closely follows that proposed in Sec.~IV of Ref.~\cite{del2024minimal}. However, our present problem differs from that considered in Ref.~\cite{del2024minimal} in that we only measure $h$ instead of $\tilde{\mathbf{S}}$ and in that we do not know the values of $r_\mathrm{A}$ and $r_\mathrm{B}$. Hence, we seek to estimate the entries of $\mathbf{S}^\mathrm{K}$ (except for the diagonal ones in the block $\mathcal{AA}$) as well as $r_\mathrm{A}$ and $r_\mathrm{B}-r_\mathrm{A}$ simultaneously. The earlier gradient-descent proposal from Ref.~\cite{sol2023experimentally} did not use an advantageous two-step procedure but it did also treat a problem with unknown load characteristics. Our algorithmic implementation consists simply in optimizing the sought-after parameters via gradient descent such that a cost function quantifying the difference between the predicted and ground-truth values of the difference of $h$ between successive configurations of the individual tunable loads is minimized. Additional implementation details can be found in Ref.~\cite{del2024minimal}. 

The algorithm is \textit{not} specific to a SISO scenario; in fact, the performance of the algorithm improves the more antennas on transmitting and receiving sides are involved, as seen in Fig.~2 in Ref.~\cite{sol2023experimentally}. We merely chose a SISO scenario in this case study because of the limited number of antennas in our experiment (see Fig.~\ref{Fig3}A), and because of the ease of defining the cost function in terms of the RSSI in the SISO case.

\subsubsection{Results}

\begin{figure}[h]
    \centering
    \includegraphics[width=\linewidth]{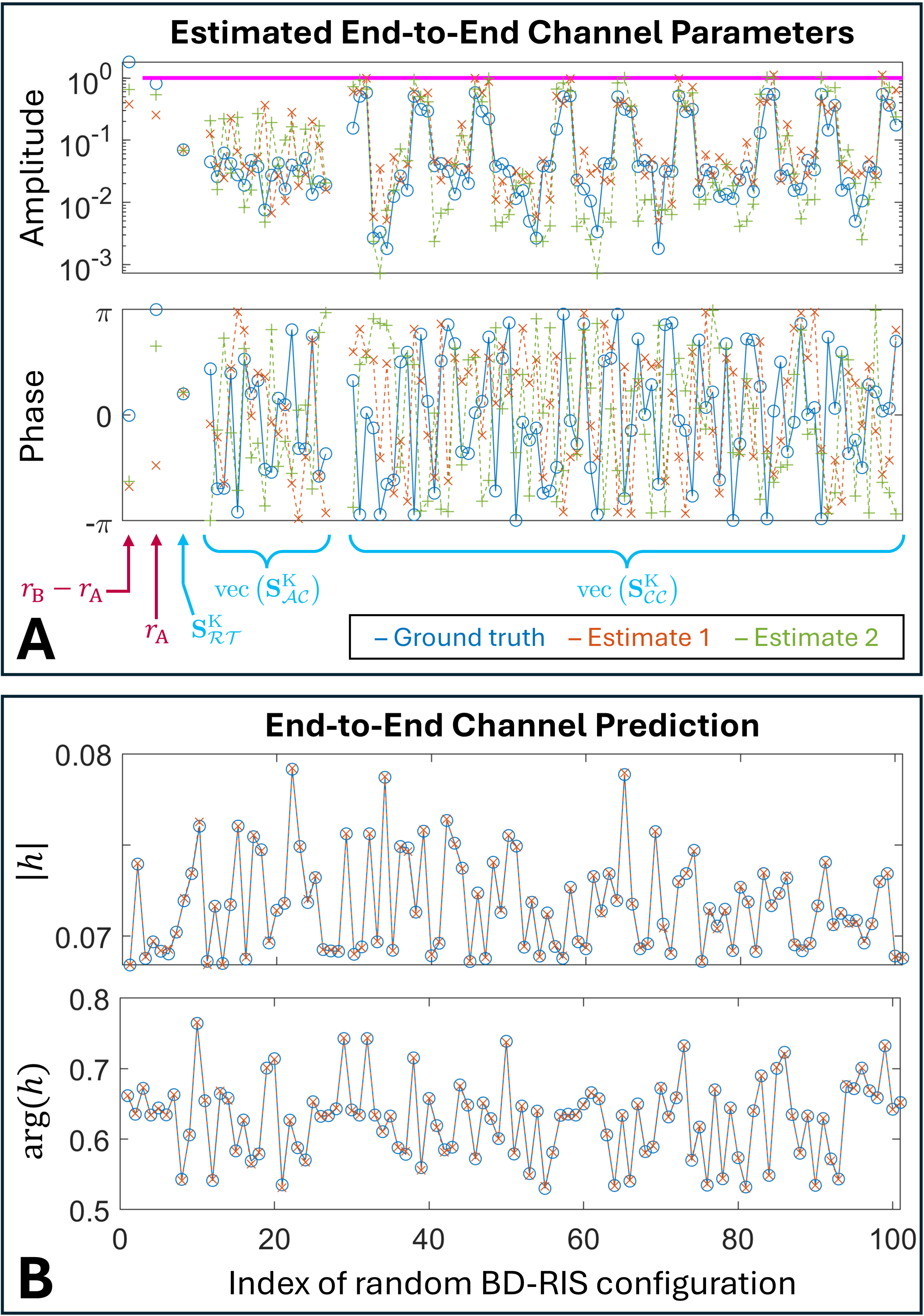}
    \caption{End-to-end BD-RIS-parametrized channel estimation in the rich-scattering environment from Fig.~\ref{Fig3}A using a physics-compliant D-RIS algorithm. The parameters to be estimated are $r_\mathrm{A}$ and $r_\mathrm{B}$, as well as the blocks $\mathcal{RT}$, $\mathcal{AC}$ and $\mathcal{CC}$ of $\mathbf{S}^\mathrm{K}$. Given these parameters, for any given $\mathbf{b}$ we can evaluate $\mathbf{S}^\mathrm{IL}$ using Eq.~(\ref{eq_b}) and then $h$ using Eq.~(\ref{eq22_new}).  (A) Amplitude and phase of the experimentally grounded ground truth (blue) and two separate estimates (red and green) of the estimated parameters. For ease of visualization, we display the matrix-valued blocks $\mathcal{AC}$ and $\mathcal{CC}$ of $\mathbf{S}^\mathrm{K}$ in vectorized form. (B) Amplitude and phase of ground-truth (blue) and predicted (red) end-to-end wireless SISO channel $h$ for 100 random unseen configurations of the BD-RIS in the rich-scattering environment from Fig.~\ref{Fig3}A.}
    \label{Fig4}
\end{figure}

Based on the same data set, we ran our end-to-end channel estimation algorithm twice. Two examples of estimated parameter sets  are plotted in Fig.~\ref{Fig4}A in red and green, and compared to the known ground-truth values plotted in blue. Except for the $\mathcal{RT}$ block of $\mathbf{S}^\mathrm{K}$ (which is a scalar in our case study), clear differences to the ground truth as well as between the two estimates are apparent, as expected due to the inevitable ambiguities. Nonetheless, the estimated set of parameters allows us to accurately predict $h(\mathbf{b})$ for arbitrary unseen $\mathbf{b}$, as evidenced by the comparison between predicted and ground-truth values of $h(\mathbf{b})$ for arbitrary unseen $\mathbf{b}$ in Fig.~\ref{Fig4}B. The inevitable ambiguities are thus not problematic for our goal of identifying an accurate mapping $\mathbf{b} \rightarrow h(\mathbf{b})$. We have hence successfully determined a ``purely physics-based digital twin'' for our BD-RIS parametrized channel without having explicitly identified $\mathbf{S}^\mathrm{RE}$ or $\mathbf{S}^\mathrm{SLC}$. In fact, our method does not even require knowledge of $N_\mathrm{S}$.

\subsection{RSSI Optimization}
\label{sec_RSSI}

Based on the estimated end-to-end channel model from the previous subsection, we can identify a configuration of the considered reconfigurable impedance network that maximizes the received signal strength indicator (RSSI, here defined as $|h|^2$) on the considered SISO link. Due to the non-linear dependence of $h$ on $\mathbf{b}$ (see also Ref.~\cite{rabault2023tacit}) and the binary constraint on the possible entries of $\mathbf{b}$, we use the simple coordinate ascent approach summarized in Algorithm~\ref{algo1} to identify an optimized configuration $\mathbf{b}_\mathrm{opt}$. This requires many forward evaluations, i.e., mappings from $\mathbf{b}$ to $h(\mathbf{b})$. However, we do not have to perform the full matrix inversion seen in Eq.~(\ref{eq22_new}) for every forward evaluation. Instead, we can use the Woodbury identity to efficiently update previous matrix inverses, as discussed in detail for the case of a D-RIS in the PhysFad formalism in Ref.~\cite{prod2023efficient}. Again, we hence benefit from the diagonal representation of the BD-RIS-parametrized channel to recycle previous algorithmic developments in the realm of D-RIS.

\begin{algorithm}[h]
Choose a random binary vector $\mathbf{b}_\mathrm{curr}\in\mathbb{C}^{N_\mathrm{C}\times 1}$.\\
Evaluate the corresponding RSSI $R_\mathrm{curr}=|h(\mathbf{b}_\mathrm{curr})|^2$.\\
$k \gets 0$.\\
\While{$k < 10N_\mathrm{C}$}{
    Define $i$ as a randomly chosen integer such that $1\leq i\leq N_\mathrm{C}$. \\
    $\mathbf{b}^\prime \gets \mathbf{b}_\mathrm{curr}$ with the $i$th entry flipped.\\
    Evaluate the corresponding RSSI $R^\prime=|h(\mathbf{b}^\prime)|^2$.\\
    \If{$R^\prime > R_\mathrm{curr}$}{
        $R_\mathrm{curr} \gets R^\prime$.\\
        $\mathbf{b}_\mathrm{curr} \gets \mathbf{b}^\prime$.
    }
    $k \gets k+1$.
}
Repeat steps 1 to 13 ten times, define the globally best $R_\mathrm{curr}$ as $R_\mathrm{opt}$, and the corresponding $\mathbf{b}_\mathrm{curr}$ as $\mathbf{b}_\mathrm{opt}$.\\
\KwOut{Optimized configuration $\mathbf{b}_{\rm opt}$ and corresponding $R_\mathrm{opt}$.}
\caption{RSSI optimization via coordinate ascent.}
\label{algo1}
\end{algorithm}

\begin{figure}[h]
    \centering
    \includegraphics[width=0.8\linewidth]{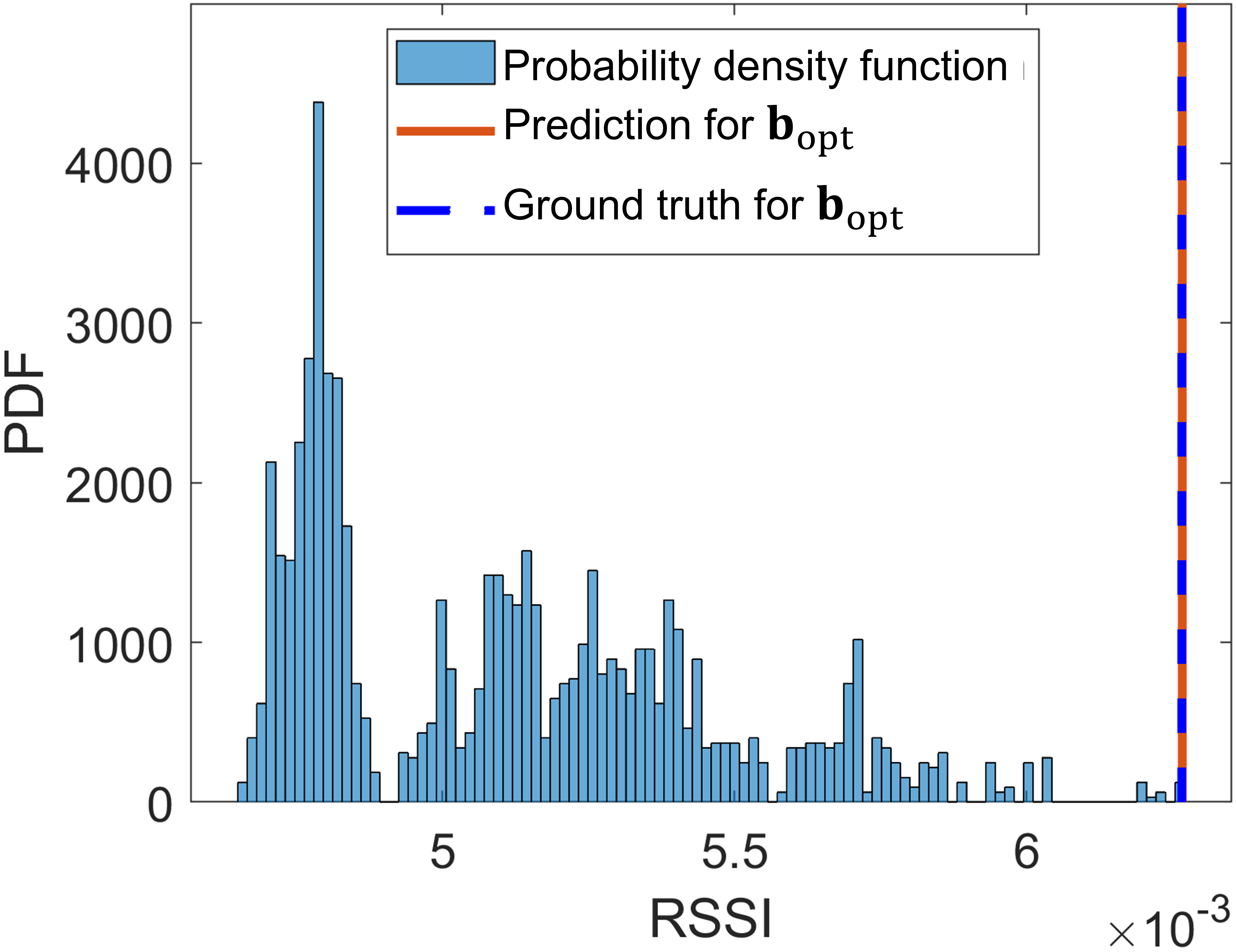}
    \caption{Probability density function of the RSSI for the considered SISO link, as well as predicted and ground-truth RSSI for the configuration $\mathbf{b}_\mathrm{opt}$ obtained with Algorithm~\ref{algo1} based on the end-to-end channel estimate from Sec.~\ref{subsec_chanest}. }
    \label{Fig5}
\end{figure}

A comparison of the expected maximal RSSI with the optimized RIS configuration $\mathbf{b}_\mathrm{opt}$ benchmarked against the ground-truth RSSI for that configuration is shown in Fig.~\ref{Fig5}. No difference is apparent upon visual inspection. For reference, the probability density function of the RSSI obtained by evaluating the RSSI for a series of random configurations is also shown. Furthermore, since an exhaustive search of the $2^9$ possible configuration is feasible for the considered small-scale case study, we also identified the globally optimal configuration (using the ground-truth simulations without reliance on the estimated channel model) which coincides with the one obtained via Algorithm~\ref{algo1}.

The RSSI enhancement enabled by the considered BD-RIS is moderate in the considered example due to the small scale of the RIS which comprises only six elements. However, this example is sufficient to demonstrate the key point of our paper, namely that the we can successfully optimize our BD-RIS based on the end-to-end channel estimate from the previous section. Both channel estimation and optimization pivotally rely on the diagonal representation of the BD-RIS-parametrized channel in Eq.~(\ref{eq22_new}).

\section{Conclusion}
\label{sec_conclusion}

To summarize, we have introduced a new \textit{diagonal} representation of BD-RIS-parametrized channels that is fully physics-compliant. Our key insight is based on the realization that the BD-RIS reconfigurable impedance network can be separated into a static part and a set of individual tunable loads. This implies that a BD-RIS-parametrized channel results from a chain cascade of radio environment (RE), static parts of the load circuit (SLC) and individual tunable loads (IL). By considering the cascade K of the static and non-diagonal systems RE and SLC, which is terminated by the tunable diagonal system IL, we achieve a representation that is directly analogous to physics-compliant models for D-RIS-parametrized channels. Therefore, our diagonal representation enables us to directly apply physics-compliant D-RIS algorithms to scenarios with BD-RIS-parametrized channels. We have evidenced this fact by using a physics-compliant D-RIS algorithm to perform end-to-end channel estimation and optimization in an experimentally grounded case study of a BD-RIS-parametrized rich-scattering radio environment.

Along the way, we have formalized the operational equivalence of physics-compliant models for D-RIS based on multi-port network theory and the coupled-dipole formalism (PhysFad). This equivalence, combined with our diagonal representation for BD-RIS within multi-port network theory, has allowed us to fill a research gap on how to describe BD-RIS-parametrized channels within PhysFad. Moreover, for pedagogical purposes, we have discussed in detail how one can determine the scattering parameters of SLC in various scenarios arising in theoretical, numerical and experimental studies. However, we have emphasized that explicitly knowing the characteristics of SLC is operationally not necessary. Indeed, our diagonal representation for BD-RIS-parametrized channels is the most compact physics-compliant representation and its deployment does not require one to explicitly separate K into RE and SLC. 

Altogether, we expect our physics-compliant diagonal representation for BD-RIS-parametrized channels to enable a paradigm shift in how practitioners in wireless communications and signal processing implement system-level optimizations of BD-RIS-parametrized radio environments. Specifically, our findings obviate the need for developing system-level optimization algorithms dedicated to BD-RIS, because existing physics-compliant algorithms already developed for D-RIS can be applied directly.

Our work also has a take-away message regarding the performance comparison between D-RIS and BD-RIS. Currently, such comparisons always fix the number of RIS elements. Our work points to the importance of comparative studies comparing for a fixed number of tunable lumped elements rather than a fixed number of RIS elements. Indeed, our work highlights that the primary entities in a BD-RIS-parametrized environment are the antennas and the tunable lumped elements in the load circuit but not the RIS elements themselves; indeed, knowledge of the number of RIS elements is not even required when using our physics-compliant diagonal representation of BD-RIS-parametrized channels, as seen in our case study.

\appendices

\section{Procedure to determine $\mathbf{S}^\mathrm{SLC,T}$ analytically}
\label{AppendixA}

In this Appendix, we illustrate the procedure for analytically determining $\mathbf{S}^\mathrm{SLC}$ for an ideal T network whose impedances have been replaced by auxiliary ports, as sketched in Fig.~\ref{Fig2}B. The characteristic impedances of the transmission lines attached to the ports are assumed to be $Z_0 = 50\ \Omega$. We follow the procedure outlined in Pozar's book~\cite{Pozar2011}: ``$S_{ij}$ is found by driving port $j$ with an incident wave of voltage $V^+_j$ and measuring the reflected wave amplitude $V^-_i$ coming out of port $i$. The incident waves on all ports except the $j$th port are set to zero, which means that all ports should be terminated in matched loads to avoid reflections.'' We also detail various sanity checks to confirm that our results are correct.

\subsection{Step-by-step procedure}

\begin{figure}[h]
    \centering
    \includegraphics[width=\linewidth]{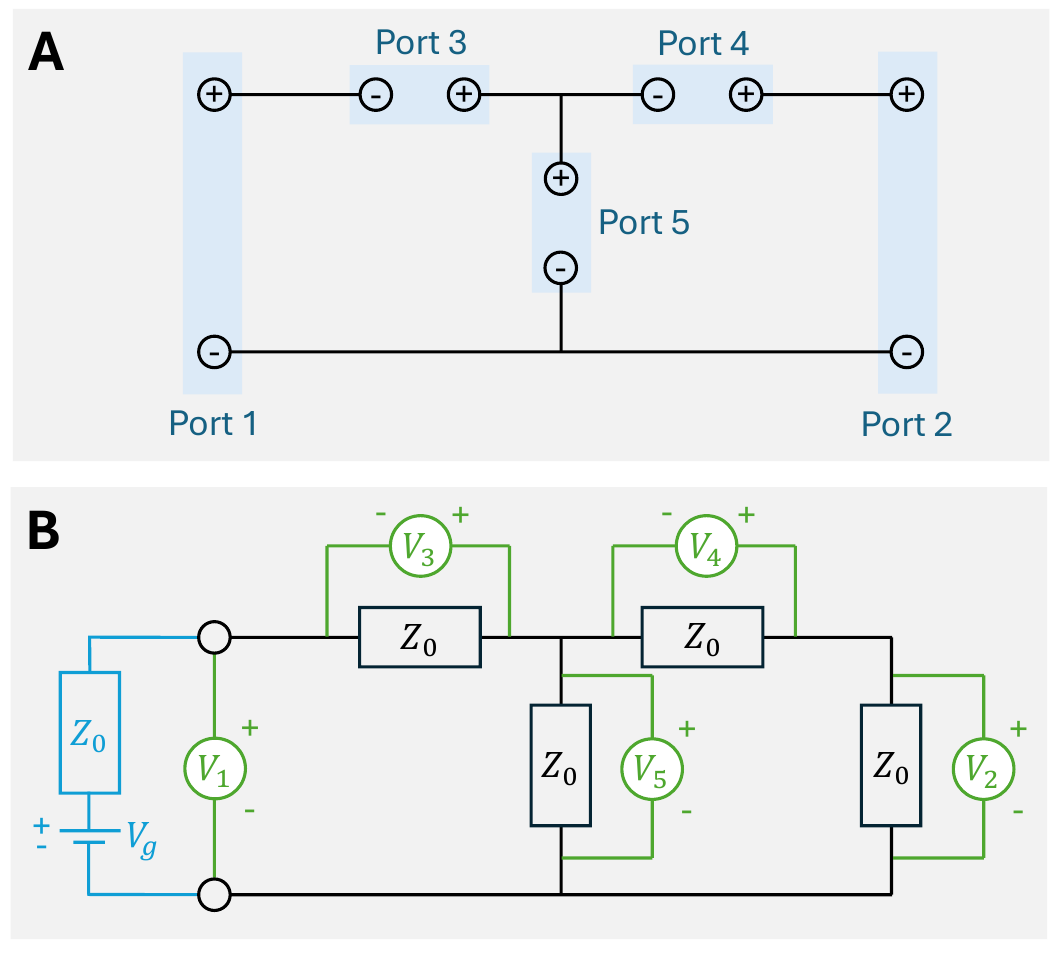}
    \caption{(A) Ideal T network with auxiliary ports replacing the three impedances. Polarities are defined for each of the five ports. (B) Setup to determine column $i=1$ of $\mathbf{S}^\mathrm{SLC}$. All ports except for the $i$th port are terminated with matched loads ($Z_0$). A matched source (voltage generator in series with a matched load, drawn in light blue) is connected to the $i$th port. The voltage drops across all ports are monitored with the voltmeters drawn in green. }
    \label{FigA1}
\end{figure}

\begin{enumerate}
    \item Define the ports, including their numbering and polarity, as done in Fig.~\ref{FigA1}A. 
    \item Replace all but the $i$th port with matched loads, and replace the $i$th port with a matched source, as done in Fig.~\ref{FigA1}B for $i=1$. A matched source is a voltage generator in series with a matched load, denoting the generator's voltage with $V_g$.
    \item Determine the voltage drop $V_j$ across the $j$th port, as illustrated using voltmeters in Fig.~\ref{FigA1}B. This is a basic circuit analysis exercise that can be performed in closed form. Alternatively, circuit analysis tools like \href{https://falstad.com/circuit/}{https://falstad.com/circuit/} can be used.
    \item Evaluate 
    \begin{equation}
        S_{ii}^\mathrm{SLC,T} = \frac{V_i^-}{V_i^+} = \frac{ V_i - V_i^+ }{ V_i^+ } = \frac{V_i - (V_g/2) }{V_g/2}.
    \end{equation}
    \item Evaluate
    \begin{equation}
        S_{ji}^\mathrm{SLC,T} = \frac{V_j^-}{V_i^+} = \frac{V_j}{V_g/2}
    \end{equation}
    for each $j\neq i$.
    \item Repeat for each $i$.    
\end{enumerate}

\subsection{Sanity checks}
For the considered problem, $\mathbf{S}^\mathrm{SLC,T}$ must satisfy two basic properties:
\begin{enumerate}
    \item \textit{Symmetry.} Due to reciprocity, we expect  
    \begin{equation}
        \left(\mathbf{S}^\mathrm{SLC,T}\right)^T = \mathbf{S}^\mathrm{SLC,T}.
    \end{equation}
    \item \textit{Unitarity.} Due to the absence of loss and gain, we expect 
    \begin{equation}
        \left(\mathbf{S}^\mathrm{SLC,T}\right)^\dagger  \mathbf{S}^\mathrm{SLC,T} = \mathbf{I}_5.
    \end{equation}    
\end{enumerate}
Verifying that these two expectations are met constitutes the first two sanity checks, which are easily passed by the form of $\mathbf{S}^\mathrm{SLC}$ defined in Eq.~(\ref{eq_T}).

Furthermore, it is well-established that the impedance matrix of an ideal T network with impedances $Z_1$, $Z_2$ and $Z_3$, as sketched in Fig.~\ref{Fig2}A, is given by
\begin{equation}
\mathbf{Z}^\mathrm{T} = \begin{bmatrix}
Z_1 + Z_3 & Z_3 \\
Z_3 & Z_2 + Z_3
\end{bmatrix} \in \mathbb{C}^{2 \times 2}.
\label{eq_Z_T}
\end{equation}
This ideal T network corresponds to terminating ports 3, 4 and 5 of our 5-port network (sketched in Fig.~\ref{Fig2}B) with load impedances $Z_1$, $Z_2$ and $Z_3$. Our third sanity check hence consists in verifying that based on the form of $\mathbf{S}^\mathrm{SLC,T}$ defined in Eq.~(\ref{eq_T}) we recover the well-established $\mathbf{Z}^\mathrm{T}$ from Eq.~(\ref{eq_Z_T}) for random complex-valued choices of the values of $Z_1$, $Z_2$ and $Z_3$. 

Since our 5-port network is characterized by a scattering matrix that cannot be converted to an impedance matrix, we first determine the reflection coefficients of the three individual load impedances using $r_i = (Z_i-Z_0)/(Z_i+Z_0)$. Then, we define $\mathbf{r} = \mathrm{diag}\left([r_1,r_2,r_3]\right)$ and evaluate
\begin{equation}
    \mathbf{S}^\mathrm{T} = \mathbf{S}^\mathrm{SLC,T}_{\bar{\mathcal{S}}\bar{\mathcal{S}}} + \mathbf{S}^\mathrm{SLC,T}_{\bar{\mathcal{S}}\mathcal{C}} \left( \left( \mathrm{diag}(\mathbf{r}) \right)^{-1} -  \mathbf{S}^\mathrm{SLC,T}_\mathcal{CC} \right)^{-1}   \mathbf{S}^\mathrm{SLC,T}_{\mathcal{C}\bar{\mathcal{S}}},
    \label{eq_load_S}
\end{equation}
where $\bar{\mathcal{S}} = \{1,2\}$ and $\mathcal{C}=\{3,4,5\}$. Finally, we confirm that $Z_0 (\mathbf{I}_2 + \mathbf{S}^\mathrm{T}) (\mathbf{I}_2 - \mathbf{S}^\mathrm{T})^{-1}$ equals $\mathbf{Z}^\mathrm{T}$ obtained via Eq.~(\ref{eq_Z_T}).

The form of $\mathbf{S}^\mathrm{SLC,T}$ defined in Eq.~(\ref{eq_T}) passes all three sanity checks.

\section{Derivation of Eq.~(\ref{eq_21})}
\label{AppendixB}

For notational ease, we use $\mathbf{\Phi} = \mathrm{diag}(\mathbf{c})$,  $\mathbf{\Sigma}=\left[ \mathbf{W}^0_{\mathcal{SS}} \right]^{-1}$, $\mathbf{\Omega}(\mathbf{c}) = \mathbf{W}^{-1}(\mathbf{c})$ and $\mathbf{\Omega}^0 = \mathbf{W}^{-1}(\mathbf{c}=\mathbf{0})$. The block-matrix inversion lemma directly yields
\begin{subequations}
\begin{equation}
    \mathbf{\Omega}^0_\mathcal{AA} = \left(\mathbf{W}^0_{\mathcal{AA}} - \mathbf{W}^0_{\mathcal{AS}} \mathbf{\Sigma} \mathbf{W}^0_{\mathcal{SA}}\right)^{-1},
    \label{eq_sigma_0_AA}
\end{equation}
\begin{equation}
     \mathbf{\Omega}^0_\mathcal{AS} = - \mathbf{\Omega}^0_\mathcal{AA} \mathbf{W}^0_{\mathcal{AS}} \mathbf{\Sigma} = \left[\mathbf{\Omega}^0_\mathcal{AS}\right]^T,
    \label{eq_sigma_0_AS}
\end{equation}
 \begin{equation}
     \mathbf{\Omega}^0_\mathcal{SS} = \mathbf{\Sigma} + \mathbf{\Sigma}\mathbf{W}^0_{\mathcal{SA}} \mathbf{\Omega}^0_\mathcal{AA} \mathbf{W}^0_{\mathcal{AS}}\mathbf{\Sigma}.
    \label{eq_sigma_0_SS}
\end{equation}   
\end{subequations}
Moreover, the Woodbury matrix identity yields the following auxiliary result:
\begin{equation}
    \left( \mathbf{W}^0_{\mathcal{SS}} - \mathbf{\Phi} \right)^{-1} = \mathbf{\Sigma} -\mathbf{\Sigma} \left( -\mathbf{\Phi} + \mathbf{\Sigma}  \right)^{-1}\mathbf{\Sigma}.
    \label{eq_aux}
\end{equation}

According to the block-matrix inversion lemma,
\begin{equation}
    \left[\mathbf{\Omega} (\mathbf{c}) \right]_\mathcal{AA} = \left[ \mathbf{W}^0_{\mathcal{AA}} - \mathbf{W}^0_{\mathcal{AS}} \left( \mathbf{W}^0_{\mathcal{SS}} - \mathbf{\Phi} \right)^{-1} \mathbf{W}^0_{\mathcal{SA}}  \right]^{-1}.
\end{equation}
Using Eq.~(\ref{eq_aux}) and Eq.~(\ref{eq_sigma_0_AA}), we obtain
\begin{equation}
    \left[\mathbf{\Omega} (\mathbf{c}) \right]_\mathcal{AA} = \Big[ \mathbf{\Omega}^0_\mathcal{AA}+ \mathbf{W}^0_{\mathcal{AS}} \mathbf{\Sigma} \left( -\mathbf{\Phi} + \mathbf{\Sigma} \right)^{-1} 
    \mathbf{\Sigma} \mathbf{W}^0_{\mathcal{SA}} \Big]^{-1}.
    \label{eq_36}
\end{equation}
Applying the Woodburry matrix identity to Eq.~(\ref{eq_36}) and using Eq.~(\ref{eq_sigma_0_AS}) and Eq.~(\ref{eq_sigma_0_SS}) yields
\begin{equation}
  \left[\mathbf{\Omega} (\mathbf{c}) \right]_\mathcal{AA} =  \mathbf{\Omega}^0_\mathcal{AA} + \mathbf{\Omega}^0_\mathcal{AS} \left( \mathbf{\Phi} - \mathbf{\Omega}^0_\mathcal{SS} \right)^{-1} \mathbf{\Omega}^0_\mathcal{SA}.  
  \label{eq32}
\end{equation}
Finally,  inserting Eq.~(\ref{eq32}) into Eq.~(\ref{eq20}) yields Eq.~(\ref{eq_21}).

\section*{Acknowledgment}
The author acknowledges stimulating discussions with L.~Le~Magoarou, M.~Nerini, H.~Prod'homme, and A.~Shaham.

\bibliographystyle{IEEEtran}

\begin{thebibliography}{10}
\providecommand{\url}[1]{#1}
\csname url@samestyle\endcsname
\providecommand{\newblock}{\relax}
\providecommand{\bibinfo}[2]{#2}
\providecommand{\BIBentrySTDinterwordspacing}{\spaceskip=0pt\relax}
\providecommand{\BIBentryALTinterwordstretchfactor}{4}
\providecommand{\BIBentryALTinterwordspacing}{\spaceskip=\fontdimen2\font plus
\BIBentryALTinterwordstretchfactor\fontdimen3\font minus \fontdimen4\font\relax}
\providecommand{\BIBforeignlanguage}[2]{{%
\expandafter\ifx\csname l@#1\endcsname\relax
\typeout{** WARNING: IEEEtran.bst: No hyphenation pattern has been}%
\typeout{** loaded for the language `#1'. Using the pattern for}%
\typeout{** the default language instead.}%
\else
\language=\csname l@#1\endcsname
\fi
#2}}
\providecommand{\BIBdecl}{\relax}
\BIBdecl

\bibitem{diagonal_BD_RIS}
P.~del Hougne, ``Physics-compliant diagonal representation of beyond-diagonal {RIS},'' \emph{Proc. SPAWC}, 2024.

\bibitem{subrt2012intelligent}
L.~Subrt and P.~Pechac, ``Intelligent walls as autonomous parts of smart indoor environments,'' \emph{IET Commun.}, vol.~6, no.~8, pp. 1004--1010, May 2012.

\bibitem{Liaskos_Visionary_2018}
C.~Liaskos, S.~Nie, A.~I. Tsioliaridou, A.~Pitsillides, S.~Ioannidis, and I.~F. Akyildiz, ``{A new wireless communication paradigm through software-controlled metasurfaces},'' \emph{IEEE Commun. Mag.}, vol.~56, no.~9, pp. 162--169, Sep. 2018.

\bibitem{del2019optimally}
P.~del Hougne, M.~Fink, and G.~Lerosey, ``Optimally diverse communication channels in disordered environments with tuned randomness,'' \emph{Nat. Electron.}, vol.~2, no.~1, pp. 36--41, 2019.

\bibitem{di2020smart}
M.~Di~Renzo, A.~Zappone, M.~Debbah, M.-S. Alouini, C.~Yuen, J.~De~Rosny, and S.~Tretyakov, ``Smart radio environments empowered by reconfigurable intelligent surfaces: How it works, state of research, and the road ahead,'' \emph{IEEE J. Sel. Areas Commun.}, vol.~38, no.~11, pp. 2450--2525, 2020.

\bibitem{alexandropoulos2021reconfigurable}
G.~C. Alexandropoulos, N.~Shlezinger, and P.~del Hougne, ``Reconfigurable intelligent surfaces for rich scattering wireless communications: Recent experiments, challenges, and opportunities,'' \emph{IEEE Commun. Mag.}, vol.~59, no.~6, pp. 28--34, 2021.

\bibitem{shen2021modeling}
S.~Shen, B.~Clerckx, and R.~Murch, ``Modeling and architecture design of reconfigurable intelligent surfaces using scattering parameter network analysis,'' \emph{IEEE Trans. Wirel. Commun.}, vol.~21, no.~2, pp. 1229--1243, 2021.

\bibitem{li2023reconfigurable}
H.~Li, S.~Shen, M.~Nerini, and B.~Clerckx, ``Reconfigurable intelligent surfaces 2.0: Beyond diagonal phase shift matrices,'' \emph{IEEE Commun. Mag.}, vol.~62, no.~3, pp. 102--108, 2024.

\bibitem{shastri2023nonlocal}
K.~Shastri and F.~Monticone, ``Nonlocal flat optics,'' \emph{Nat. Photonics}, vol.~17, no.~1, pp. 36--47, 2023.

\bibitem{chen2023nonlocal}
Y.~Chen, M.~A. Abouelatta, K.~Wang, M.~Kadic, and M.~Wegener, ``Nonlocal cable-network metamaterials,'' \emph{Adv. Mater.}, vol.~35, no.~15, p. 2209988, 2023.

\bibitem{sol2024covert}
J.~Sol, M.~R{\"o}ntgen, and P.~del Hougne, ``Covert scattering control in metamaterials with non-locally encoded hidden symmetry,'' \emph{Adv. Mater.}, vol.~36, no.~11, p. 2303891, 2024.

\bibitem{sol2022meta}
J.~Sol, D.~R. Smith, and P.~del Hougne, ``Meta-programmable analog differentiator,'' \emph{Nat. Commun.}, vol.~13, no.~1, p. 1713, 2022.

\bibitem{sol2023reflectionless}
J.~Sol, A.~Alhulaymi, A.~D. Stone, and P.~del Hougne, ``Reflectionless programmable signal routers,'' \emph{Sci. Adv.}, vol.~9, no.~4, p. eadf0323, 2023.

\bibitem{faul2024agile}
F.~T. Faul, L.~Cronier, A.~Alhulaymi, A.~D. Stone, and P.~del Hougne, ``Agile free-form signal filtering with a chaotic-cavity-backed non-local programmable metasurface,'' \emph{arXiv:2407.00054}, 2024.

\bibitem{denicke2012application}
E.~Denicke, M.~Henning, H.~Rabe, and B.~Geck, ``The application of multiport theory for {MIMO RFID} backscatter channel measurements,'' \emph{Proc. EuMC}, pp. 522--525, 2012.

\bibitem{V2NA_2p0}
P.~del Hougne, ``Virtual {VNA} 2.0: Ambiguity-free scattering matrix estimation by terminating inaccessible ports with tunable and coupled loads,'' \emph{arXiv:2409.10977}, 2024.

\bibitem{faqiri2022physfad}
R.~Faqiri, C.~Saigre-Tardif, G.~C. Alexandropoulos, N.~Shlezinger, M.~F. Imani, and P.~del Hougne, ``{PhysFad}: Physics-based end-to-end channel modeling of {RIS}-parametrized environments with adjustable fading,'' \emph{IEEE Trans. Wirel. Commun.}, vol.~22, no.~1, pp. 580--595, 2022.

\bibitem{prod2023efficient}
H.~Prod’homme and P.~del Hougne, ``Efficient computation of physics-compliant channel realizations for (rich-scattering) {RIS}-parametrized radio environments,'' \emph{IEEE Commun. Lett.}, vol.~27, no.~12, pp. 3375--3379, 2023.

\bibitem{sol2023experimentally}
J.~Sol, H.~Prod’homme, L.~Le~Magoarou, and P.~del Hougne, ``Experimentally realized physical-model-based frugal wave control in metasurface-programmable complex media,'' \emph{Nat. Commun.}, vol.~15, no.~1, p. 2841, 2024.

\bibitem{rabault2023tacit}
A.~Rabault, L.~Le~Magoarou, J.~Sol, G.~C. Alexandropoulos, N.~Shlezinger, H.~V. Poor, and P.~del Hougne, ``On the tacit linearity assumption in common cascaded models of {RIS}-parametrized wireless channels,'' \emph{IEEE Trans. Wirel. Commun.}, vol.~23, no.~8, pp. 10\,001--10\,014, 2024.

\bibitem{seguinot1998multimode}
C.~Seguinot, P.~Kennis, J.-F. Legier, F.~Huret, E.~Paleczny, and L.~Hayden, ``Multimode {TRL}. a new concept in microwave measurements: theory and experimental verification,'' \emph{IEEE Trans. Microw. Theory Techn.}, vol.~46, no.~5, pp. 536--542, 1998.

\bibitem{ivrlavc2010toward}
M.~T. Ivrla{\v{c}} and J.~A. Nossek, ``Toward a circuit theory of communication,'' \emph{IEEE Trans. Circuits Syst. I: Regul. Pap.}, vol.~57, no.~7, pp. 1663--1683, 2010.

\bibitem{ivrlavc2014multiport}
------, ``The multiport communication theory,'' \emph{IEEE Circuits Syst. Mag.}, vol.~14, no.~3, pp. 27--44, 2014.

\bibitem{gradoni_EndtoEnd_2020}
G.~Gradoni and M.~Di~Renzo, ``End-to-end mutual coupling aware communication model for reconfigurable intelligent surfaces: An electromagnetic-compliant approach based on mutual impedances,'' \emph{IEEE Wirel. Commun. Lett.}, vol.~10, no.~5, pp. 938--942, 2021.

\bibitem{tap2022}
Z.~Zhang, J.~W. Zhang, J.~W. Wu, J.~C. Liang, Z.~X. Wang, Q.~Cheng, Q.~S. Cheng, T.~J. Cui, H.~Q. Yang, G.~B. Liu, and S.~R. Wang, ``Macromodeling of reconfigurable intelligent surface based on microwave network theory,'' \emph{IEEE Trans. Antennas Propag.}, vol.~70, no.~10, pp. 8707--8717, 2022.

\bibitem{badheka2023accurate}
D.~Badheka, J.~Sapis, S.~R. Khosravirad, and H.~Viswanathan, ``Accurate modeling of intelligent reflecting surface for communication systems,'' \emph{IEEE Trans. Wirel. Commun.}, vol.~22, no.~9, pp. 5871--5883, 2023.

\bibitem{mursia2023}
P.~Mursia, S.~Phang, V.~Sciancalepore, G.~Gradoni, and M.~D. Renzo, ``{SARIS}: Scattering aware reconfigurable intelligent surface model and optimization for complex propagation channels,'' \emph{IEEE Wirel. Commun. Lett.}, vol.~12, no.~11, pp. 1921--1925, 2023.

\bibitem{tapie2023systematic}
J.~Tapie, H.~Prod’homme, M.~F. Imani, and P.~del Hougne, ``Systematic physics-compliant analysis of over-the-air channel equalization in {RIS}-parametrized wireless networks-on-chip,'' \emph{IEEE J. Sel. Areas Commun.}, vol.~42, no.~8, pp. 2026--2038, 2024.

\bibitem{Akrout_asilomar_2023}
M.~Akrout, F.~Bellili, A.~Mezghani, and J.~A. Nossek, ``Physically consistent models for intelligent reflective surface-assisted communications under mutual coupling and element size constraint,'' \emph{Proc. ACSSC}, pp. 1589--1594, 2023.

\bibitem{del2024minimal}
P.~del Hougne, ``Virtual {VNA}: Minimal-ambiguity scattering matrix estimation with load-tunable ports,'' \emph{arXiv:2403.08074}, 2024.

\bibitem{nossek2024}
J.~A. Nossek, D.~Semmler, M.~Joham, and W.~Utschick, ``Physically consistent modeling of wireless links with reconfigurable intelligent surfaces using multiport network analysis,'' \emph{IEEE Wirel. Commun. Lett.}, vol.~13, no.~8, pp. 2240--2244, 2024.

\bibitem{konno2024}
K.~Konno, S.~Terranova, Q.~Chen, and G.~Gradoni, ``Generalised impedance model of wireless links assisted by reconfigurable intelligent surfaces,'' \emph{IEEE Trans. Antennas Propag.}, vol.~72, no.~10, pp. 7691--7699, 2024.

\bibitem{nerini2024universal}
M.~Nerini, S.~Shen, H.~Li, M.~D. Renzo, and B.~Clerckx, ``A universal framework for multiport network analysis of reconfigurable intelligent surfaces,'' \emph{IEEE Trans. Wirel. Commun.}, vol.~23, no.~10, pp. 14\,575--14\,590, 2024.

\bibitem{franek_eucap_2024}
O.~Franek, ``Electromagnetics-based {RIS} channel model with near-field accuracy improvement,'' \emph{Proc. EuCAP}, pp. 1--5, 2024.

\bibitem{viikari2024}
A.~D. Kuznetsov, J.~Holopainen, and V.~Viikari, ``Predicting the bistatic scattering of a multiport loaded structure under arbitrary excitation: The {S}-parameters approach,'' \emph{IEEE Trans. Antennas Propag.}, vol.~72, no.~8, pp. 6691--6701, 2024.

\bibitem{abrardo2024}
A.~Abrardo, A.~Toccafondi, and M.~Di~Renzo, ``Design of reconfigurable intelligent surfaces by using s-parameter multiport network theory – optimization and full-wave validation,'' \emph{IEEE Trans. Wirel. Commun.}, vol.~23, no.~11, pp. 17\,084--17\,102, 2024.

\bibitem{BDRIS_renzo_clerckx_2024}
H.~Li, S.~Shen, M.~Nerini, M.~Di~Renzo, and B.~Clerckx, ``Beyond diagonal reconfigurable intelligent surfaces with mutual coupling: Modeling and optimization,'' \emph{IEEE Commun. Lett.}, vol.~28, no.~4, pp. 937--941, 2024.

\bibitem{king1949measurement}
D.~King, ``The measurement and interpretation of antenna scattering,'' \emph{Proc. IRE}, vol.~37, no.~7, pp. 770--777, 1949.

\bibitem{hansen1989relationships}
R.~C. Hansen, ``Relationships between antennas as scatterers and as radiators,'' \emph{Proc. IEEE}, vol.~77, no.~5, pp. 659--662, 1989.

\bibitem{hansen1990antenna}
R.~Hansen, ``Antenna mode and structural mode {RCS}: dipole,'' \emph{Microw. Opt. Technol. Lett.}, vol.~3, no.~1, pp. 6--10, 1990.

\bibitem{anderson_cascade_1966}
B.~D.~O. Anderson and R.~W. Newcomb, ``\BIBforeignlanguage{en}{Cascade connection for time-invariant n-port networks},'' \emph{\BIBforeignlanguage{en}{Proc. Inst. Electr. Eng.}}, vol. 113, no.~6, pp. 970--974, Jun. 1966.

\bibitem{ha1981solid}
T.~T. Ha, \emph{Solid-state microwave amplifier design}.\hskip 1em plus 0.5em minus 0.4em\relax Wiley-Interscience, 1981.

\bibitem{ferrero1992new}
A.~Ferrero, U.~Pisani, and K.~J. Kerwin, ``A new implementation of a multiport automatic network analyzer,'' \emph{IEEE Trans. Microw. Theory Tech.}, vol.~40, no.~11, pp. 2078--2085, 1992.

\bibitem{prod2024efficient}
H.~Prod’homme and P.~del Hougne, ``Efficient and updatable evaluation of arbitrarily complex connections between multi-port networks,'' \emph{arXiv:2412.17884}, 2024.

\bibitem{redheffer_inequalities_1959}
R.~Redheffer, ``Inequalities for a {Matrix} {Riccati} {Equation},'' \emph{J. Math. Mech.}, vol.~8, no.~3, pp. 349--367, 1959.

\bibitem{chu_generalized_1986}
T.~S. Chu and T.~Itoh, ``Generalized scattering matrix method for analysis of cascaded and offset microstrip step discontinuities,'' \emph{IEEE Trans. Microw. Theory Tech.}, vol.~34, no.~2, pp. 280--284, 1986.

\bibitem{overfelt1989alternate}
P.~Overfelt and D.~White, ``Alternate forms of the generalized composite scattering matrix,'' \emph{IEEE Trans. Microw. Theory Tech.}, vol.~37, no.~8, pp. 1267--1268, 1989.

\bibitem{sol2024optimal}
J.~Sol, L.~Le~Magoarou, and P.~del Hougne, ``Optimal blind focusing on perturbation-inducing targets in sub-unitary complex media,'' \emph{Laser Photonics Rev.}, p. 2400619, 2024.

\bibitem{murch2016}
S.~Shen and R.~D. Murch, ``Impedance matching for compact multiple antenna systems in random {RF} fields,'' \emph{IEEE Trans. Antennas Propag.}, vol.~64, no.~2, pp. 820--825, 2016.

\bibitem{macom}
\BIBentryALTinterwordspacing
MACOM. {MADP-000907-14020x}. [Online]. Available: \url{https://cdn.macom.com/datasheets/MADP-000907-14020x.pdf}
\BIBentrySTDinterwordspacing

\bibitem{li2024channel}
H.~Li, S.~Shen, Y.~Zhang, and B.~Clerckx, ``Channel estimation and beamforming for beyond diagonal reconfigurable intelligent surfaces,'' \emph{arXiv:2403.18087}, 2024.

\bibitem{de2024channel}
A.~L. de~Almeida, B.~Sokal, H.~Li, and B.~Clerckx, ``Channel estimation for beyond diagonal {RIS} via tensor decomposition,'' \emph{arXiv:2407.20402}, 2024.

\bibitem{Pozar2011}
D.~M. Pozar, \emph{Microwave Engineering}, 4th~ed.\hskip 1em plus 0.5em minus 0.4em\relax Hoboken, NJ: Wiley, 2011.

\end{thebibliography}

\providecommand{\noopsort}[1]{}\providecommand{\singleletter}[1]{#1}%

\end{document}